\documentclass[preprint,showpacs,showkeys,amsmath,amssymb]{revtex4}

\usepackage{graphicx}

\allowdisplaybreaks

\begin{document}

\title{Size of Isospin Breaking in Charged $K_{\ell 4}$ Decay}

\author{A.~Nehme}
\email{miryama.nehme@wanadoo.fr}
\affiliation{
27 rue du Four de la Terre \\
F-84000 Avignon, France}

\date{\today}

\begin{abstract}
We evaluate the size of isospin breaking corrections to form factors
$f$ and $g$ of the $K_{\ell 4}$ decay process
$K^+\rightarrow\pi^+\pi^-\ell^+\nu_{\ell}$ which is actually
measured by the extended NA48 setup at CERN. We found that, keeping
apart the effect of Coulomb interaction, isospin breaking does not
affect modules. This is due to the cancelation between corrections
of electromagnetic origin and those generated by the difference
between up and down quark masses. On the other hand,
electromagnetism affects considerably phases if the infrared
divergence is dropped out using a minimal subtraction scheme.
Consequently, the greatest care must be taken in the extraction of
$\pi\pi$ phase shifts from experiment.
\end{abstract}


\maketitle

\section{Introduction}

Measuring the quark condensate remains the main concern for
physicists of non perturbative quantum chromodynamics. The purest
process allowing a direct measurement of this parameter is $\pi\pi$
scattering. Information concerning the latter can be obtained from
the rescattering of two pions in the final state of
pionium~\cite{Knecht:2002gz}, $K\rightarrow
3\pi$~\cite{Nehme:2004xf}, or $K_{\ell 4}$~\cite{Nehme:2003bz}
decays. Let $\delta_l^I$ be the phase of a two-pion state of angular
momentum $l$ and isospin $I$ and consider the charged $K_{\ell 4}$
decay process
\begin{equation} \label{eq:process}
K^+(p)\rightarrow\pi^+(p_1)\pi^-(p_2)\ell^+(p_{\ell})\nu_{\ell}(p_{\nu})\,,
\end{equation}
where the lepton $\ell$ is either a muon $\mu$ or an electron $e$,
and $\nu$ stands for the corresponding neutrino. In the isospin
limit, the decay amplitude $\mathcal{A}$ for process
(\ref{eq:process}) can be parameterized in terms of three vectorial
($F$, $G$, and $R$) and one anomalous ($H$) form factors:
\begin{eqnarray}
\mathcal{A} &\doteq&
\frac{i}{\sqrt{2}}\,G_FV_{us}^*\overline{u}(\boldsymbol{p}_{\nu})\gamma_{\mu}(1-\gamma^5)v(\boldsymbol{p}_{\ell})\times
\nonumber \\
&& \left\lbrace \dfrac{i}{M_{K^{\pm}}}\left[
(p_1+p_2)^{\mu}F+(p_1-p_2)^{\mu}G+(p_{\ell}+p_{\nu})^{\mu}R\right]
\right.
\nonumber \\
&& \left. -\dfrac{1}{M_{K^{\pm}}^3}\,
\epsilon^{\mu\nu\rho\sigma}(p_{\ell}+p_{\nu})_{\nu}
(p_1+p_2)_{\rho}(p_1-p_2)_{\sigma}H\right\rbrace \,,
\end{eqnarray}
where $V_{us}$ denotes the Cabibbo-Kobayashi-Maskawa flavor-mixing
matrix element and $G_F$ is the so-called Fermi coupling constant.
Note that form factors are made dimensionless by inserting the
normalizations, $M_{K^{\pm}}^{-1}$ and $M_{K^{\pm}}^{-3}$. The fact
that we have used the \textit{charged} kaon mass is a purely
conventional matter and corresponds to the choice of defining the
Isospin limit in terms of charged masses.

Form factors are analytic functions of three independent Lorentz
invariants,
\begin{equation}
s_{\pi}\,\doteq\,(p_1+p_2)^2\,, \qquad
s_{\ell}\,\doteq\,(p_{\ell}+p_{\nu})^2\,,
\end{equation}
and the angle $\theta_{\pi}$ formed by $\boldsymbol{p}_1$, in the
dipion rest frame, and the line of flight of the dipion as defined
in the kaon rest frame~\cite{Cabibbo:1965, Cabibbo:1965E}. In the
following, we will be interested only in two form factors, $F$ and
$G$, and consider the partial wave expansion,
\begin{eqnarray}
F &=&
\widetilde{f}_S(s_{\pi},s_{\ell})e^{i\delta_0^0(s_{\pi})}+\widetilde{f}_P(s_{\pi},s_{\ell})\cos\theta_{\pi}e^{i\delta_1^1(s_{\pi})}\,,
\\
G &=&
\widetilde{g}_P(s_{\pi},s_{\ell})e^{i\delta_1^1(s_{\pi})}+\widetilde{g}_D(s_{\pi},s_{\ell})\cos\theta_{\pi}e^{i\delta_2^0(s_{\pi})}\,,
\end{eqnarray}
where a convenient parametrization of $\widetilde{f}_S$,
$\widetilde{f}_P$, $\widetilde{g}_P$, and $\widetilde{g}_D$ in the
experimentally relevant region has been proposed in
Ref.~\cite{Amoros:1999mg}.

The currently running NA48 experiment aims at measuring form factors
for $K_{\ell 4}$ decay of the charged kaon with an accuracy better
than the one offered by previous
measurement~\cite{Pislak:2001bf,Pislak:2003sv}. The outgoing data on
form factors contain, besides a strong interaction contribution, a
contribution coming from the electroweak interaction. The latter
breaks isospin symmetry and is expected to be sizable near the
$\pi\pi$ production threshold~\cite{Stern:private}. In order to
extract $\pi\pi$ scattering parameters from the NA48 measurement,
the isospin breaking correction to form factors should therefore be
under control. In this direction, we recently published analytic
expressions for $F$ and $G$ form factors calculated at one-loop
level in the framework of chiral perturbation theory based on the
effective Lagrangian including mesons, photons, and
leptons~\cite{Cuplov:2003bj}. In the present work, we will use the
method proposed in Ref.~\cite{Nehme:2004ui} to split analytically
the isospin limit and isospin breaking part in form factors,
allowing a first evaluation of isospin breaking effects in charged
$K_{\ell 4}$ decays.

\section{A brief review of the method}

We shall start things off by the general form of the decay amplitude
for process (\ref{eq:process}) as dictated by Lorentz covariance,
\begin{eqnarray}
{\cal A} &\doteq & \frac{G_FV_{us}^*}{\sqrt{2}}\,{\overline
u}(\boldsymbol{p}_{\nu})
        (1+\gamma^5)\times
\nonumber \\
&& \left\lbrace \dfrac{1}{M_{K^{\pm}}}\left[
(p_1+p_2)^{\mu}f+(p_1-p_2)^{\mu}g+(p_{\ell}+p_{\nu})^{\mu}r\right]
\gamma_{\mu} \right.
\nonumber \\
&& +\dfrac{i}{M_{K^{\pm}}^3}\,
\epsilon^{\mu\nu\rho\sigma}(p_{\ell}+p_{\nu})_{\nu}
(p_1+p_2)_{\rho}(p_1-p_2)_{\sigma}h
\nonumber \\
&& \left.
+\frac{1}{2M_{K^{\pm}}^2}\,[\gamma_{\mu},\gamma_{\nu}]\,p_1^{\mu}p_2^{\nu}\,T\right\rbrace
v(\boldsymbol{p}_l)\,. \nonumber
\end{eqnarray}
The quantities $f$, $g$, $r$, and $h$, will be called the
\textit{corrected} $K_{\ell 4}$ form factors since their isospin
limits are nothing else than the $K_{\ell 4}$ form factors, $F$,
$G$, $R$, and $H$, respectively. The tensorial form factor $T$ is
purely isospin breaking and has been calculated at leading chiral
order in Ref.~\cite{Cuplov:2003bj}. The corrected form factors as
well as the tensorial one are analytic functions of five independent
Lorentz invariants, $s_{\pi}$, $s_{\ell}$, $\theta_{\pi}$,
$\theta_{\ell}$, and $\phi$. $\theta_{\ell}$ is the angle formed by
$\boldsymbol{p}_{\ell}$, in the dilepton rest frame, and the line of
flight of the dilepton as defined in the kaon rest frame. $\phi$ is
the angle between the normals to the planes defined in the kaon rest
frame by the pion pair and the lepton pair, respectively. Let us
denote by $\delta F$ and $\delta G$ the next-to-leading order
corrections to the $F$ and $G$ form factors, respectively,
\begin{eqnarray}
f &=& \frac{M_{K^{\pm}}}{\sqrt{2}F_0}\,\bigg (\,1+\delta F\,\bigg
)\,,
\nonumber \\
g &=& \frac{M_{K^{\pm}}}{\sqrt{2}F_0}\,\bigg (\,1+\delta G\,\bigg
)\,. \nonumber
\end{eqnarray}
The analytic expressions for $\delta F$ and $\delta G$ were given
in~\cite{Cuplov:2003bj}. We shall distinguish between
\textit{photonic} and \textit{non photonic} contributions to $\delta
F$ and $\delta G$. The photonic contribution comes from those
Feynman diagrams with a virtual photon exchanged between two meson
legs or one meson leg and a pure strong vertex. Obviously, this
contribution is proportional to $e^2$, where $e$ is the electric
charge, and depends in general on the five independent kinematical
variables, $s_{\pi}$, $s_{\ell}$, $\theta_{\pi}$, $\theta_{\ell}$,
and $\phi$ through Lorentz invariants like $(p_2+p_{\ell})^2$, say.
The non photonic contribution comes from diagrams having similar
topology as the ones in the pure strong theory with Isospin breaking
allowed in propagators and vertices. This contribution generates
Isospin breaking terms proportional to the rate of $SU(2)$ to
$SU(3)$ breaking,
\begin{equation}
\epsilon\,\doteq\,\dfrac{\sqrt{3}}{4}\,\dfrac{m_d-m_u}{m_s-\hat{m}}\,,
\qquad \hat{m}\,\doteq\,\frac{1}{2}\,(m_u+m_d)\,,
\end{equation}
and to mass square difference between charged and neutral mesons,
\begin{eqnarray}
\Delta_{\pi} &\doteq&
M_{\pi^{\pm}}^2-M_{\pi^0}^2\,=\,2Z_0e^2F_0^2+\mathcal{O}(p^4)\,,
\\
\Delta_K &\doteq&
M_{K^{\pm}}^2-M_{K^0}^2\,=\,2Z_0e^2F_0^2-B_0(m_d-m_u)+\mathcal{O}(p^4)\,,
\end{eqnarray}
or equivalently, $(m_d-m_u)/(m_s-\hat{m})$, $Z_0e^2$, and $m_d-m_u$.
The kinematical dependence is on three Lorentz invariants,
$(p_1+p_2)^2$, $(p-p_1)^2$, and $(p-p_2)^2$ which represent
respectively the dipion mass square, the exchange energy between the
kaon and the neutral pion, and that between the kaon and the charged
pion. In terms of independent kinematical variables, the preceding
scalars are functions of $s_{\pi}$, $s_{\ell}$, and
$\cos\theta_{\pi}$.

It has been noted in Ref.~\cite{Nehme:2004ui} that for
\begin{equation} \label{eq:assumption}
s_{\ell}\,=\,m_{\ell}^2
\end{equation}
the photonic contribution neither depends on $\theta_{\ell}$ nor on
$\phi$ and, consequently, it can be written as
\begin{equation}
\label{eq:photonic_expansion}
\mathrm{photonic~contribution}\,=\,e^2\,\varsigma
(s_{\pi})+e^2\,\vartheta (s_{\pi})\cos\theta_{\pi} \,,
\end{equation}
where $\varsigma$ and $\vartheta$ are analytic functions of
$s_{\pi}$. Note that, to the order we are working, that is, to
leading order in isospin breaking, the power counting scheme we use
dictates the following on-shell conditions to be used in the
argument of $\varsigma$ and $\vartheta$,
\begin{equation}
\label{eq:on-shell_conditions}
p^2\,=\,M_K^2\,\doteq\,B_0(m_s+\hat{m})\,, \quad
p_1^2\,=\,p_2^2\,=\,M_{\pi}^2\,\doteq\,2B_0\hat{m}\,.
\end{equation}

With respect to the nonphotonic contribution, it depends on
$s_{\pi}$, $(p-p_1)^2$, $(p-p_2)^2$ and masses through one- and
two-point functions. In order to split strong and electromagnetic
interactions in one-point functions we use the formula
\begin{equation}
\label{eq:splitting_one} A(M_{P^0}^2)\,=\,A(M_{P^{\pm}}^2)+\left[
\dfrac{1}{16\pi^2}-\dfrac{1}{M_P^2}\,A(M_P^2)\right] \Delta_P\,,
\end{equation}
where $P$ denotes a pion, $\pi$, or a kaon, $K$, and $\Delta_P$ the
difference,
\begin{equation}
\label{eq:difference} \Delta_P\,\doteq\,M_{P^{\pm}}^2-M_{P^0}^2\,.
\end{equation}
Concerning the splitting in two-point functions $B(p_1,m_0,m_1)$, we
have to expand exchange energies in powers of the fine structure
constant $\alpha$ and $m_d-m_u$. We then inject the obtained
expansion in the expression of $B(p_1^2+\delta
,m_0^2+\delta_0,m_1^2+\delta_1)$ where $\delta$, $\delta_0$, and
$\delta_1$, are leading order in isospin breaking,
\begin{equation}
\delta\,,\,\delta_0\,,\,\delta_1\,,\,=\,\mathcal{O}(\alpha\,,\,m_d-m_u)\,.
\end{equation}
The final step consists on expanding two-point functions to first
order in $\delta$, $\delta_0$, and $\delta_1$,
\begin{eqnarray}
\label{eq:splitting_two} && B(p_1^2+\delta
,m_0^2+\delta_0,m_1^2+\delta_1)\,=\,B(p_1^2,m_0^2,m_1^2)
\nonumber \\
&& -\dfrac{1}{32\pi^2p_1^2}\left[ \ln\left(
\dfrac{m_0^2}{m_1^2}\right) +(p_1^2+m_1^2-m_0^2)\tau
(p_1^2,m_0^2,m_1^2)\right] \delta_0
\nonumber \\
&& +\dfrac{1}{32\pi^2p_1^2}\left[ \ln\left(
\dfrac{m_0^2}{m_1^2}\right) -(p_1^2-m_1^2+m_0^2)\tau
(p_1^2,m_0^2,m_1^2)\right] \delta_1
\nonumber \\
&& -\dfrac{1}{32\pi^2p_1^4}\left\lbrace
2p_1^2+(m_1^2-m_0^2)\ln\left( \dfrac{m_0^2}{m_1^2}\right) \right.
\nonumber \\
&& +\left[ (m_1^2-m_0^2)^2-p_1^2(m_1^2+m_0^2)\right] \tau
(p_1^2,m_0^2,m_1^2)\,\bigg\}\,\delta\,,
\end{eqnarray}
with $\tau$ a generic integral defined by,
\begin{equation}
\label{eq:tau_definition} \tau (p_1^2,m_0^2,m_1^2)\,\doteq\,
\int_0^1dx\dfrac{1}{xm_0^2+(1-x)m_1^2-x(1-x)p_1^2}\,.
\end{equation}

Putting all this together, form factors for $K_{\ell 4}$ decay of
the charged kaon can be written in the following compact form which
shows explicitly the splitting between strong and electromagnetic
interactions,
\begin{eqnarray}
\label{eq:principal_one} && x\left(
s_{\pi},(p-p_1)^2,(p-p_2)^2,(p_2+p_{\ell})^2,\ldots \right) \,=\,
\nonumber \\
&& \qquad \dfrac{M_{K^{\pm}}}{\sqrt{2}F_0}\left[
1+U^x(s_{\pi})+V^x(s_{\pi})\cos\theta_{\pi}\right] \,, \;\;
x\,=\,f\,,\; g\,,
\end{eqnarray}
where,
\begin{eqnarray}
\label{eq:principal_two} W^x &=&
W_{\mathrm{s}}^x+W_{\pi}^x\Delta_{\pi}+W_K^x\Delta_K
\nonumber \\
& & \qquad
+W_{e^2}^xe^2+W_{\epsilon}^x\,\dfrac{\epsilon}{\sqrt{3}}\,, \quad
W\,=\,U\,,\; V\,,
\end{eqnarray}
are analytic functions of $s_{\pi}$. If one makes the following
substitutions,
\begin{eqnarray}
\Delta_{\pi} &\longrightarrow & 2Z_0e^2F_0^2\,,
\label{eq:difference_pi}
\\
\Delta_K &\longrightarrow &
2Z_0e^2F_0^2-\dfrac{4\epsilon}{\sqrt{3}}\,(M_K^2-M_{\pi}^2)\,,
\label{eq:difference_K}
\end{eqnarray}
then, equations (\ref{eq:principal_one}) and
(\ref{eq:principal_two}) read,
\begin{eqnarray}
W^x &=&
W_{\mathrm{s}}^x+W_{\alpha}^xe^2+W_{m_d-m_u}^x\,\dfrac{\epsilon}{\sqrt{3}}\,,
\label{eq:main_one}
\\
W_{\alpha}^x &=& W_{e^2}^x+2Z_0F_0^2(W_{\pi}^x+W_K^x)\,,
\label{eq:main_two}
\\
W_{m_d-m_u}^x &=& W_{\epsilon}^x-4(M_K^2-M_{\pi}^2)W_K^x\,.
\label{eq:main_three}
\end{eqnarray}
The aim of the present work is to determine the $U$ functions
corresponding to $f$ and $g$ form factors for $K_{\ell 4}$ decay of
the charged kaon.

\section{Isospin limit}

We have
\begin{eqnarray}
U_s^f &=& -\frac{1}{384\pi^2F_0^2}\left[ 20M_{K^{\pm}}^2+ 7
M_{\pi^{\pm}}^2+9M_{\eta}^2 \right. \nonumber\\
&& \left. -6t_{\pi}+\frac{3}{t_{\pi}}\left( 2M_{K^{\pm}}^2+
M_{\pi^{\pm}}^2+M_{\eta}^2\right)(M_{\pi^{\pm}}^2-M_{K^{\pm}}^2)\right]
\nonumber\\
&& +\frac{2}{F_0^2}\left[
16(s_{\pi}-2M_{\pi^{\pm}}^2)L_1+4(M_{K^{\pm}}^2-m_{\ell}^2+s_{\pi})L_2
\right. \nonumber\\
&& \left.
+(M_{K^{\pm}}^2-8M_{\pi^{\pm}}^2-m_{\ell}^2+5s_{\pi})L_3-2(2M_{K^{\pm}}^2-7M_{\pi^{\pm}}^2)L_4+m_{\ell}^2L_9\right]
\nonumber\\
&& -\frac{1}{8F_0^2}\left[ 5-\frac{2}{t_{\pi}}\left(
M_{K^{\pm}}^2-2M_{\pi^{\pm}}^2\right)+\frac{2}{t_{\pi}^2}\left(
M_{\pi^{\pm}}^2-M_{K^{\pm}}^2\right)^2\right] A(M_{\pi^{\pm}}^2)
\nonumber\\
&& +\frac{1}{8F_0^2}\left[ 2-\frac{1}{t_{\pi}}\left(
8M_{K^{\pm}}^2-5M_{\pi^{\pm}}^2-3M_{\eta}^2\right) \right.
\nonumber\\
&& \left. -\frac{2}{t_{\pi}^2}\left(
2M_{K^{\pm}}^2-M_{\pi^{\pm}}^2-M_{\eta}^2\right)
(M_{\pi^{\pm}}^2-M_{K^{\pm}}^2)\right] A(M_{K^{\pm}}^2) \nonumber\\
&& +\frac{1}{8F_0^2}\left[ 3+\frac{1}{t_{\pi}}\left(
2M_{K^{\pm}}^2+3M_{\pi^{\pm}}^2-3M_{\eta}^2\right) \right.
\nonumber\\
&& \left. -\frac{2}{t_{\pi}^2}\left(
M_{\pi^{\pm}}^2-M_{K^{\pm}}^2\right)
(M_{\eta}^2-M_{K^{\pm}}^2)\right] A(M_{\eta}^2) \nonumber\\
&& +\frac{1}{12F_0^2}\left[
6(2s_{\pi}-M_{\pi^{\pm}}^2)B(s_{\pi},M_{\pi^{\pm}}^2,M_{\pi^{\pm}}^2)
\right. \nonumber\\
&& \left.
+6M_{\pi^{\pm}}^2B(s_{\pi},M_{\eta}^2,M_{\eta}^2)+9s_{\pi}B(s_{\pi},M_{K^{\pm}}^2,M_{K^{\pm}}^2)\right]
\nonumber\\
&& +\frac{1}{4F_0^2}\left[ 3M_{K^{\pm}}^2+M_{\pi^{\pm}}^2-4t_{\pi}
\right. \nonumber\\
&& \left. +\frac{2}{t_{\pi}}\left(
M_{\pi^{\pm}}^2-M_{K^{\pm}}^2\right)^2+\frac{1}{t_{\pi}^2}\left(
M_{\pi^{\pm}}^2-M_{K^{\pm}}^2\right)^3\right]
B(t_{\pi},M_{\pi^{\pm}}^2,M_{K^{\pm}}^2) \nonumber\\
&& -\frac{1}{8F_0^2}\left[
2M_{K^{\pm}}^2-M_{\pi^{\pm}}^2+3M_{\eta}^2-\frac{1}{t_{\pi}}\left(
4M_{K^{\pm}}^2+M_{\pi^{\pm}}^2-5M_{\eta}^2\right) M_{K^{\pm}}^2
\right. \nonumber\\
&& \left. +\frac{3}{t_{\pi}}\left( M_{\pi^{\pm}}^2-M_{\eta}^2\right)
M_{\eta}^2-\frac{2}{t_{\pi}^2}\left(
M_{\pi^{\pm}}^2-M_{K^{\pm}}^2\right)
(M_{\eta}^2-M_{K^{\pm}}^2)^2\right]
B(t_{\pi},M_{\eta}^2,M_{K^{\pm}}^2)\,, \\
U_s^g &=& -\frac{1}{384\pi^2F_0^2}\left[
12M_{K^{\pm}}^2+21M_{\pi^{\pm}}^2+3M_{\eta}^2 \right. \nonumber\\
&& \left. -4s_{\pi}-2t_{\pi}-\frac{3}{t_{\pi}}\left(
2M_{K^{\pm}}^2+M_{\pi^{\pm}}^2+M_{\eta}^2\right)
(M_{\pi^{\pm}}^2-M_{K^{\pm}}^2)\right] \nonumber\\
&& -\frac{2}{F_0^2}\left[
(M_{K^{\pm}}^2-m_{\ell}^2+s_{\pi})L_3+2(M_{\pi^{\pm}}^2+2M_{K^{\pm}}^2)L_4-m_{\ell}^2L_9\right]
\nonumber\\
&& -\frac{1}{24F_0^2}\left[
5-\frac{6}{t_{\pi}}\,M_{\pi^{\pm}}^2-\frac{6}{t_{\pi}^2}\left(
M_{\pi^{\pm}}^2-M_{K^{\pm}}^2\right)^2\right] A(M_{\pi^{\pm}}^2)
\nonumber\\
&& +\frac{1}{24F_0^2}\left[ 2+\frac{3}{t_{\pi}}\left(
4M_{K^{\pm}}^2-3M_{\pi^{\pm}}^2-M_{\eta}^2\right) \right.
\nonumber\\
&& \left. +\frac{6}{t_{\pi}^2}\left(
2M_{K^{\pm}}^2-M_{\pi^{\pm}}^2-M_{\eta}^2\right)
(M_{\pi^{\pm}}^2-M_{K^{\pm}}^2)\right] A(M_{K^{\pm}}^2) \nonumber\\
&& +\frac{1}{8F_0^2}\left[ 1+\frac{1}{t_{\pi}}\left(
M_{\eta}^2-3M_{\pi^{\pm}}^2\right) +\frac{2}{t_{\pi}^2}\left(
M_{\pi^{\pm}}^2-M_{K^{\pm}}^2\right)
(M_{\eta}^2-M_{K^{\pm}}^2)\right] A(M_{\eta}^2) \nonumber\\
&& +\frac{1}{12F_0^2}\left[
2(s_{\pi}-4M_{\pi^{\pm}}^2)B(s_{\pi},M_{\pi^{\pm}}^2,M_{\pi^{\pm}}^2)+(s_{\pi}-4M_{K^{\pm}}^2)B(s_{\pi},M_{K^{\pm}}^2,M_{K^{\pm}}^2)\right]
\nonumber\\
&& -\frac{1}{4F_0^2}\left[ M_{K^{\pm}}^2-M_{\pi^{\pm}}^2-t_{\pi}
\right. \nonumber\\
&& \left. +\frac{1}{t_{\pi}}\left(
M_{\pi^{\pm}}^2-M_{K^{\pm}}^2\right)^2+\frac{1}{t_{\pi}^2}\left(
M_{\pi^{\pm}}^2-M_{K^{\pm}}^2\right)^3\right]
B(t_{\pi},M_{\pi^{\pm}}^2,M_{K^{\pm}}^2) \nonumber\\
&& -\frac{1}{8F_0^2}\left[
2M_{K^{\pm}}^2+M_{\pi^{\pm}}^2+M_{\eta}^2-2t_{\pi} \right.
\nonumber\\
&& +\frac{1}{t_{\pi}}\left(
2M_{K^{\pm}}^2+M_{\pi^{\pm}}^2-3M_{\eta}^2\right)
M_{K^{\pm}}^2+\frac{1}{t_{\pi}}\left(
2M_{K^{\pm}}^2-3M_{\pi^{\pm}}^2+M_{\eta}^2\right) M_{\eta}^2
\nonumber\\
&& \left. +\frac{2}{t_{\pi}^2}\left(
M_{\pi^{\pm}}^2-M_{K^{\pm}}^2\right)
(M_{\eta}^2-M_{K^{\pm}}^2)^2\right]
B(t_{\pi},M_{\eta}^2,M_{K^{\pm}}^2)\,.
\end{eqnarray}
In the preceding expressions, we used the notation
\begin{equation}
t_{\pi}\,=\,\frac{1}{2}\,(M_{K^{\pm}}^2+2M_{\pi^{\pm}}^2+m_{\ell}^2-s_{\pi})\,.
\end{equation}
Note that in the isospin breaking correction, the same expression
holds for $t_{\pi}$ with the replacement $M_{P^{\pm}}\rightarrow
M_P$.

\section{non-photonic correction}

The correction due to $\epsilon$ reads:
\begin{eqnarray}
U_{\epsilon}^f &=& \frac{1}{4F_0^2}\left\{
(M_{\pi}^2-M_{\eta}^2)\left[ 3+\frac{1}{t_{\pi}}\left(
M_{\pi}^2-M_K^2\right)\right]\frac{1}{16\pi^2} \right. \nonumber\\
&& +\left[ -3+\frac{2}{t_{\pi}}\,M_{\pi}^2+\frac{2}{t_{\pi}^2}\left(
M_{\pi}^2-M_K^2\right)^2\right] A(M_{\pi}^2) -\frac{2}{t_{\pi}^2}\left( M_{\pi}^2-M_{\eta}^2\right) (M_{\pi}^2-M_K^2)A(M_K^2) \nonumber\\
&& +\left[ 3-\frac{2}{t_{\pi}}\,M_{\pi}^2-\frac{2}{t_{\pi}^2}\left(
M_{\pi}^2-M_K^2\right) (M_{\eta}^2-M_K^2)\right]
A(M_{\eta}^2)+12(s_{\pi}-M_{\pi}^2)B(s_{\pi},M_{\pi}^2,M_{\pi}^2)
\nonumber\\
&&
-4M_{\pi}^2B(s_{\pi},M_{\eta}^2,M_{\eta}^2)-4(3s_{\pi}-4M_{\pi}^2)B(s_{\pi},M_{\pi}^2,M_{\eta}^2)
\nonumber\\
&& +2\left[ M_K^2+5M_{\pi}^2-3t_{\pi}-\frac{1}{t_{\pi}}\left(
M_{\pi}^2-M_K^2\right)^2-\frac{1}{t_{\pi}^2}\left(
M_{\pi}^2-M_K^2\right)^3\right] B(t_{\pi},M_{\pi}^2,M_K^2)
\nonumber\\
&& +\left[ -2(5M_K^2+M_{\pi}^2-3t_{\pi})+\frac{1}{t_{\pi}}\left(
3M_{\eta}^2-5M_K^2\right) (M_{\pi}^2-M_K^2) \right. \nonumber\\
&& \left.\left. +\frac{1}{t_{\pi}}\left( 3M_{\eta}^2-M_K^2\right)
(M_{\eta}^2-M_K^2)+\frac{2}{t_{\pi}^2}\left( M_{\pi}^2-M_K^2\right)
(M_{\eta}^2-M_K^2)^2\right] B(t_{\pi},M_{\eta}^2,M_K^2)\right\}\,,
\\
U_{\epsilon}^g &=& -\frac{1}{4F_0^2}\left\{
(M_{\pi}^2-M_{\eta}^2)\left[ -1+\frac{1}{t_{\pi}}\left(
M_{\pi}^2-M_K^2\right)\right]\frac{1}{16\pi^2} \right. \nonumber\\
&& +\left[ 1+\frac{2}{t_{\pi}}\,M_K^2+\frac{2}{t_{\pi}^2}\left(
M_{\pi}^2-M_K^2\right)^2\right] A(M_{\pi}^2) \nonumber\\
&& -\frac{1}{t_{\pi}}\left[
4M_K^2-3M_{\pi}^2-M_{\eta}^2+\frac{2}{t_{\pi}}\left(
M_{\pi}^2-M_{\eta}^2\right) (M_{\pi}^2-M_K^2)\right] A(M_K^2)
\nonumber\\
&& +\left[ -1+\frac{1}{t_{\pi}}\left(
2M_K^2-3M_{\pi}^2-M_{\eta}^2\right) -\frac{2}{t_{\pi}^2}\left(
M_{\pi}^2-M_K^2\right) (M_{\eta}^2-M_K^2)\right] A(M_{\eta}^2)
\nonumber\\
&& -2\left[ M_K^2-3M_{\pi}^2+2t_{\pi}+\frac{1}{t_{\pi}^2}\left(
M_{\pi}^2-M_K^2\right)^3\right] B(t_{\pi},M_{\pi}^2,M_K^2)
\nonumber\\
&& +\left[
-2(3M_K^2+M_{\pi}^2-2M_{\eta}^2-2t_{\pi})-\frac{1}{t_{\pi}}\left(
4M_K^2-3M_{\eta}^2\right) (M_{\pi}^2-M_K^2) \right. \nonumber\\
&& \left.\left. +\frac{1}{t_{\pi}}\left( 4M_K^2+M_{\eta}^2\right)
(M_{\eta}^2-M_K^2)+\frac{2}{t_{\pi}^2}\left( M_{\pi}^2-M_K^2\right)
(M_{\eta}^2-M_K^2)^2\right] B(t_{\pi},M_{\eta}^2,M_K^2)\right\}\,.
\end{eqnarray}
That due to $M_{\pi^{\pm}}^2-M_{\pi^0}^2$ is equal to:
\begin{eqnarray}
U_{\pi}^f &=& -\frac{1}{384\pi^2F_0^2}\left[
6-\frac{1}{t_{\pi}}\left( M_K^2-7M_{\pi}^2\right)
+\frac{2}{t_{\pi}^2}\left( M_{\pi}^2-M_K^2\right)^2\right]
-\frac{24}{F_0^2}\,L_4 \nonumber\\
&& +\frac{1}{24F_0^2}\left[ 9+\frac{2}{t_{\pi}}\left(
13M_{\pi}^2-M_K^2\right) -\frac{2}{t_{\pi}^2}\left(
M_K^2-2M_{\pi}^2\right)
(M_{\pi}^2-M_K^2)\right]\frac{A(M_{\pi}^2)}{M_{\pi}^2} \nonumber\\
&& -\frac{1}{24F_0^2}\,\frac{1}{t_{\pi}}\left[
9+\frac{2}{t_{\pi}}\left( M_{\pi}^2-M_K^2\right)\right]
A(M_K^2)-\frac{3}{8F_0^2}\,\frac{1}{t_{\pi}}\,A(M_{\eta}^2)
\nonumber\\
&& -\frac{1}{384\pi^2F_0^2}\left[ 6-\frac{1}{t_{\pi}}\left(
5M_K^2+M_{\pi}^2\right) \right. \nonumber\\
&& \left. +\frac{2}{t_{\pi}^2}\left( M_K^2-2M_{\pi}^2\right)
(M_{\pi}^2-M_K^2)-\frac{1}{t_{\pi}^3}\left(
M_{\pi}^2-M_K^2\right)^3\right]\ln\left(\frac{M_{\pi}^2}{M_K^2}\right)
\nonumber\\
&& -\frac{1}{384\pi^2F_0^2}\left[
M_K^2-7M_{\pi}^2+6t_{\pi}+\frac{1}{t_{\pi}}\left(
7M_K^2-3M_{\pi}^2\right) (M_{\pi}^2-M_K^2) \right. \nonumber\\
&& \left. -\frac{1}{t_{\pi}^2}\left( M_K^2-3M_{\pi}^2\right)
(M_{\pi}^2-M_K^2)^2+\frac{1}{t_{\pi}^3}\left(
M_{\pi}^2-M_K^2\right)^4\right]\tau (t_{\pi},M_{\pi}^2,M_K^2)
\nonumber\\
&& +\frac{1}{32\pi^2F_0^2}\left( s_{\pi}-M_{\pi}^2\right)\tau
(s_{\pi},M_{\pi}^2,M_{\pi}^2)+\frac{5}{2F_0^2}\,B(s_{\pi},M_{\pi}^2,M_{\pi}^2)
\nonumber\\
&& -\frac{1}{2F_0^2}\left[
B(s_{\pi},M_{\eta}^2,M_{\eta}^2)-4B(s_{\pi},M_K^2,M_K^2)\right]
\nonumber\\
&& +\frac{1}{12F_0^2}\left[ 3+\frac{1}{t_{\pi}}\left(
10M_K^2-13M_{\pi}^2\right) -\frac{2}{t_{\pi}^2}\left(
M_{\pi}^2-M_K^2\right)^2\right] B(t_{\pi},M_{\pi}^2,M_K^2) \nonumber\\
&& -\frac{3}{8F_0^2}\left[ 1-\frac{1}{t_{\pi}}\left(
M_{\eta}^2-M_K^2\right)\right] B(t_{\pi},M_{\eta}^2,M_K^2)\,, \\
U_{\pi}^g &=& -\frac{1}{384\pi^2F_0^2}\left[
10-\frac{1}{t_{\pi}}\left( M_K^2+5M_{\pi}^2\right)
-\frac{2}{t_{\pi}^2}\left( M_{\pi}^2-M_K^2\right)^2\right]
+\frac{8}{F_0^2}\,L_4 \nonumber\\
&& +\frac{1}{24F_0^2}\left[
11-\frac{22}{t_{\pi}}\,M_{\pi}^2+\frac{2}{t_{\pi}^2}\left(
M_K^2-2M_{\pi}^2\right)
(M_{\pi}^2-M_K^2)\right]\frac{A(M_{\pi}^2)}{M_{\pi}^2} \nonumber\\
&& +\frac{1}{24F_0^2}\,\frac{1}{t_{\pi}}\left[
7+\frac{2}{t_{\pi}}\left( M_{\pi}^2-M_K^2\right)\right]
A(M_K^2)+\frac{3}{8F_0^2}\,\frac{1}{t_{\pi}}\,A(M_{\eta}^2)
\nonumber\\
&& +\frac{1}{384\pi^2F_0^2}\left[ 7-\frac{1}{t_{\pi}}\left(
7M_K^2+3M_{\pi}^2\right) \right. \nonumber\\
&& \left. +\frac{1}{t_{\pi}^2}\left( M_K^2-3M_{\pi}^2\right)
(M_{\pi}^2-M_K^2)-\frac{1}{t_{\pi}^3}\left(
M_{\pi}^2-M_K^2\right)^3\right]\ln\left(\frac{M_{\pi}^2}{M_K^2}\right)
\nonumber\\
&& -\frac{1}{384\pi^2F_0^2}\left[
10M_{\pi}^2-7t_{\pi}-\frac{8}{t_{\pi}}\,M_K^2(M_{\pi}^2-M_K^2)
\right. \nonumber\\
&& \left.
-\frac{2}{t_{\pi}^2}\,M_{\pi}^2(M_{\pi}^2-M_K^2)^2-\frac{1}{t_{\pi}^3}\left(
M_{\pi}^2-M_K^2\right)^4\right]\tau (t_{\pi},M_{\pi}^2,M_K^2)
\nonumber\\
&& +\frac{1}{12F_0^2}\left[ 23+\frac{1}{t_{\pi}}\left(
11M_{\pi}^2-8M_K^2\right) +\frac{2}{t_{\pi}^2}\left(
M_{\pi}^2-M_K^2\right)^2\right] B(t_{\pi},M_{\pi}^2,M_K^2)
\nonumber\\
&& +\frac{3}{8F_0^2}\left[ 1-\frac{1}{t_{\pi}}\left(
M_{\eta}^2-M_K^2\right)\right] B(t_{\pi},M_{\eta}^2,M_K^2)\,.
\end{eqnarray}
The one due to $M_{K^{\pm}}^2-M_{K^0}^2$ is given by:
\begin{eqnarray}
U_K^f &=& \frac{1}{384\pi^2F_0^2}\left[ 18-\frac{1}{t_{\pi}}\left(
20M_K^2-11M_{\pi}^2-9M_{\eta}^2\right)\right] \nonumber\\
&&
+\frac{4}{F_0^2}\,L_4-\frac{1}{12F_0^2}\,\frac{1}{t_{\pi}^2}\left(
M_{\pi}^2-M_K^2\right) A(M_{\pi}^2) \nonumber\\
&& -\frac{1}{24F_0^2}\left[ 6-\frac{1}{t_{\pi}}\left( 22
M_K^2-7M_{\pi}^2-9M_{\eta}^2\right) \right. \nonumber\\
&& \left. -\frac{2}{t_{\pi}^2}\left(
8M_K^2-M_{\pi}^2-3M_{\eta}^2\right)
(M_{\pi}^2-M_K^2)\right]\frac{A(M_K^2)}{M_K^2} \nonumber\\
&& -\frac{1}{4F_0^2}\,\frac{1}{t_{\pi}}\left[
1+\frac{1}{t_{\pi}}\left( M_{\pi}^2-M_K^2\right)\right]
A(M_{\eta}^2) \nonumber\\
&& +\frac{1}{384\pi^2F_0^2}\left[ 6-\frac{1}{t_{\pi}}\left(
5M_K^2+M_{\pi}^2\right) \right. \nonumber\\
&& \left. +\frac{2}{t_{\pi}^2}\left( M_K^2-2M_{\pi}^2\right)
(M_{\pi}^2-M_K^2)-\frac{1}{t_{\pi}^3}\left(
M_{\pi}^2-M_K^2\right)^3\right]\ln\left(\frac{M_{\pi}^2}{M_K^2}\right)
\nonumber\\
&& +\frac{1}{256\pi^2F_0^2}\,\frac{1}{t_{\pi}}\left[
2M_K^2-M_{\pi}^2+3M_{\eta}^2+\frac{1}{t_{\pi}}\left(
5M_{\eta}^2-M_{\pi}^2-4M_K^2\right) M_K^2 \right. \nonumber\\
&& \left. +\frac{3}{t_{\pi}}\left( M_{\pi}^2-M_{\eta}^2\right)
M_{\eta}^2-\frac{2}{t_{\pi}^2}\left( M_{\pi}^2-M_K^2\right)
(M_{\eta}^2-M_K^2)^2\right]\ln\left(\frac{M_{\eta}^2}{M_K^2}\right)
\nonumber\\
&& +\frac{1}{64\pi^2F_0^2}\,s_{\pi}\tau (s_{\pi},M_K^2,M_K^2)
\nonumber\\
&& +\frac{1}{384\pi^2F_0^2}\left[
11M_K^2-5M_{\pi}^2-6t_{\pi}+\frac{1}{t_{\pi}}\left(
3M_K^2+5M_{\pi}^2\right) (M_{\pi}^2-M_K^2) \right. \nonumber\\
&& \left. -\frac{1}{t_{\pi}^2}\left( 3M_K^2-5M_{\pi}^2\right)
(M_{\pi}^2-M_K^2)^2+\frac{1}{t_{\pi}^3}\left(
M_{\pi}^2-M_K^2\right)^4\right]\tau (t_{\pi},M_{\pi}^2,M_K^2)
\nonumber\\
&& -\frac{1}{256\pi^2F_0^2}\left[
2M_K^2-M_{\pi}^2+3M_{\eta}^2-\frac{6}{t_{\pi}}\left(
M_{\eta}^2-M_K^2\right)^2-\frac{1}{t_{\pi}^2}\left(
6M_K^2-M_{\pi}^2-5M_{\eta}^2\right) M_K^2 \right. \nonumber\\
&& \left. +\frac{1}{t_{\pi}^2}\left(
2M_K^2+M_{\pi}^2-3M_{\eta}^2\right)
M_{\eta}^2-\frac{2}{t_{\pi}^3}\left( M_{\pi}^2-M_K^2\right)
(M_{\eta}^2-M_K^2)^3\right]\tau (t_{\pi},M_{\eta}^2,M_K^2)
\nonumber\\
&& -\frac{1}{12F_0^2}\left[ 3+\frac{1}{t_{\pi}}\left(
2M_K^2-5M_{\pi}^2\right) -\frac{2}{t_{\pi}^2}\left(
M_{\pi}^2-M_K^2\right)^2\right] B(t_{\pi},M_{\pi}^2,M_K^2) \nonumber\\
&& +\frac{1}{8F_0^2}\left[ 4-\frac{1}{t_{\pi}}\left(
6M_K^2+M_{\pi}^2-5M_{\eta}^2\right) +\frac{4}{t_{\pi}^2}\left(
M_{\pi}^2-M_K^2\right) (M_{\eta}^2-M_K^2)\right]
B(t_{\pi},M_{\eta}^2,M_K^2)\,, \\
U_K^g &=& -\frac{1}{384\pi^2F_0^2}\left[ 14-\frac{3}{t_{\pi}}\left(
4M_K^2-3M_{\pi}^2-M_{\eta}^2\right)\right] +\frac{4}{F_0^2}\,L_4
\nonumber\\
&& -\frac{1}{12F_0^2}\,\frac{1}{t_{\pi}}\left[
1-\frac{1}{t_{\pi}}\left( M_{\pi}^2-M_K^2\right)\right]
A(M_{\pi}^2)+\frac{1}{4F_0^2}\,\frac{1}{t_{\pi}^2}\left(
M_{\pi}^2-M_K^2\right) A(M_{\eta}^2) \nonumber\\
&& +\frac{1}{24F_0^2}\left[ 10-\frac{1}{t_{\pi}}\left(
6M_K^2-5M_{\pi}^2-3M_{\eta}^2\right) \right. \nonumber\\
&& \left. -\frac{2}{t_{\pi}^2}\left(
8M_K^2-M_{\pi}^2-3M_{\eta}^2\right)
(M_{\pi}^2-M_K^2)\right]\frac{A(M_K^2)}{M_K^2} \nonumber\\
&& -\frac{1}{384\pi^2F_0^2}\left[ 7-\frac{1}{t_{\pi}}\left(
7M_K^2+3M_{\pi}^2\right) \right. \nonumber\\
&& \left. +\frac{1}{t_{\pi}^2}\left( M_K^2-3M_{\pi}^2\right)
(M_{\pi}^2-M_K^2)-\frac{1}{t_{\pi}^3}\left(
M_{\pi}^2-M_K^2\right)^3\right]\ln\left(\frac{M_{\pi}^2}{M_K^2}\right)
\nonumber\\
&& -\frac{1}{256\pi^2F_0^2}\left[ 2-\frac{1}{t_{\pi}}\left(
2M_K^2+M_{\pi}^2+M_{\eta}^2\right) -\frac{1}{t_{\pi}^2}\left(
2M_K^2+M_{\pi}^2-3M_{\eta}^2\right) M_K^2 \right. \nonumber\\
&& \left. -\frac{1}{t_{\pi}^2}\left(
2M_K^2-3M_{\pi}^2+M_{\eta}^2\right)
M_{\eta}^2-\frac{2}{t_{\pi}^3}\left( M_{\pi}^2-M_K^2\right)
(M_{\eta}^2-M_K^2)^2\right]\ln\left(\frac{M_{\eta}^2}{M_K^2}\right)
\nonumber\\
&& -\frac{1}{192\pi^2F_0^2}\left( s_{\pi}-4M_K^2\right)\tau
(s_{\pi},M_K^2,M_K^2) \nonumber\\
&& -\frac{1}{384\pi^2F_0^2}\left[
14M_K^2-4M_{\pi}^2-7t_{\pi}+\frac{6}{t_{\pi}}\left(
M_{\pi}^2+M_K^2\right) (M_{\pi}^2-M_K^2) \right. \nonumber\\
&& \left. -\frac{2}{t_{\pi}^2}\left( M_K^2-2M_{\pi}^2\right)
(M_{\pi}^2-M_K^2)^2+\frac{1}{t_{\pi}^3}\left(
M_{\pi}^2-M_K^2\right)^4\right]\tau (t_{\pi},M_{\pi}^2,M_K^2)
\nonumber\\
&& -\frac{1}{256\pi^2F_0^2}\left[
4M_K^2+M_{\pi}^2-M_{\eta}^2-2t_{\pi}-\frac{2}{t_{\pi}}\left(
M_{\pi}^2-M_{\eta}^2\right) M_{\eta}^2 \right. \nonumber\\
&& -\frac{1}{t_{\pi}^2}\left( M_{\pi}^2-M_{\eta}^2\right)
(M_{\eta}^2-M_K^2)M_{\eta}^2+\frac{1}{t_{\pi}^2}\left(
4M_K^2-M_{\pi}^2-3M_{\eta}^2\right) (M_{\eta}^2-M_K^2)M_K^2
\nonumber\\
&& \left. +\frac{2}{t_{\pi}^3}\left( M_{\pi}^2-M_K^2\right)
(M_{\eta}^2-M_K^2)^3\right]\tau
(t_{\pi},M_{\eta}^2,M_K^2)-\frac{1}{3F_0^2}\,B(s_{\pi},M_K^2,M_K^2)
\nonumber\\
&& +\frac{1}{12F_0^2}\left[
5-\frac{3}{t_{\pi}}\,M_{\pi}^2-\frac{2}{t_{\pi}^2}\left(
M_{\pi}^2-M_K^2\right)^2\right] B(t_{\pi},M_{\pi}^2,M_K^2)
\nonumber\\
&& +\frac{1}{8F_0^2}\,\frac{1}{t_{\pi}}\left[
2M_K^2+M_{\pi}^2-M_{\eta}^2-\frac{4}{t_{\pi}}\left(
M_{\pi}^2-M_K^2\right) (M_{\eta}^2-M_K^2)\right]
B(t_{\pi},M_{\eta}^2,M_K^2)\,.
\end{eqnarray}

\section{photonic correction}

We have
\begin{eqnarray}
U_{e^2}^f &=& -\frac{1}{32\pi^2}\left[
9+2\ln\left(\frac{m_{\gamma}^2}{m_{\ell}^2}\right)
+4\ln\left(\frac{m_{\gamma}^2}{M_{\pi}^2}\right)
+2\ln\left(\frac{m_{\gamma}^2}{M_K^2}\right)\right] \nonumber\\
&& -\frac{1}{18}\left(
24K_1-264K_2-16K_5-88K_6-36K_{12}+120X_1+9X_6\right) \nonumber\\
&&
-\frac{1}{2}\,\frac{A(m_{\ell}^2)}{m_{\ell}^2}+2\,\frac{A(M_{\pi}^2)}{M_{\pi}^2}+\frac{A(M_K^2)}{M_K^2}-\frac{1}{2t_{\pi}}\left[
A(M_{\pi}^2)-A(M_K^2)\right] \nonumber\\
&& -\frac{m_{\ell}^2}{2}\left\{\frac{3}{t_{\pi}}-\frac{1}{\lambda
(t_{\pi},m_{\ell}^2,M_{\pi}^2)}\left[
-\frac{m_{\ell}^2}{t_{\pi}}\left(
M_K^2+5M_{\pi}^2-3m_{\ell}^2\right)\right.\right. \nonumber\\
&& \left.\left. +\frac{M_{\pi}^2}{t_{\pi}}\left(
M_K^2+2M_{\pi}^2\right)
+M_K^2-12M_{\pi}^2-5m_{\ell}^2+2t_{\pi}\right]\right\}
B(0,m_{\ell}^2,M_K^2) \nonumber\\
&&
+\frac{m_{\ell}^2}{2}\,\frac{M_K^2-2M_{\pi}^2+m_{\ell}^2-2t_{\pi}}{\lambda
(t_{\pi},m_{\ell}^2,M_{\pi}^2)}\left[
1+\frac{12m_{\ell}^2M_{\pi}^2}{\lambda
(t_{\pi},m_{\ell}^2,M_{\pi}^2)}\right] B(m_{\ell}^2,0,M_K^2)
\nonumber\\
&& +\left[ 2+\frac{m_{\ell}^2}{\lambda
(s_{\pi},m_{\ell}^2,M_K^2)}\left(
M_K^2-3m_{\ell}^2+3s_{\pi}\right)\right. \nonumber\\
&& \left. +\frac{3m_{\ell}^2}{\lambda
(t_{\pi},m_{\ell}^2,M_{\pi}^2)}\left(
M_{\pi}^2+m_{\ell}^2-t_{\pi}\right)\right]
B(m_{\ell}^2,0,m_{\ell}^2) \nonumber\\
&& +\left\{ 2-\frac{8M_{\pi}^2(M_{\pi}^2-t_{\pi})}{\lambda
(t_{\pi},M_{\pi}^2,M_K^2)}-\frac{12m_{\ell}^2M_{\pi}^2}{\lambda
(t_{\pi},m_{\ell}^2,M_{\pi}^2)}\right. \nonumber\\
&&
+\frac{3m_{\ell}^2M_{\pi}^2}{\lambda^2(t_{\pi},m_{\ell}^2,M_{\pi}^2)}\left[
-M_{\pi}^2(M_K^2-4M_{\pi}^2)-m_{\ell}^2(M_K^2+3M_{\pi}^2-m_{\ell}^2)
\right. \nonumber\\
&& \left.\left. +(M_K^2-4M_{\pi}^2-m_{\ell}^2)t_{\pi}\right]\right\}
B(M_{\pi}^2,0,M_{\pi}^2) \nonumber\\
&& +\left[ 1-\frac{2M_K^2}{\lambda (s_{\pi},m_{\ell}^2,M_K^2)}\left(
M_K^2-2m_{\ell}^2-s_{\pi}\right)\right. \nonumber\\
&& \left.
-\frac{4M_K^2}{\lambda (t_{\pi},M_{\pi}^2,M_K^2)}\left(
M_K^2-3M_{\pi}^2-t_{\pi}\right)\right] B(M_K^2,0,M_K^2) \nonumber\\
&& +\left\{ -2+\frac{1}{\lambda (s_{\pi},m_{\ell}^2,M_K^2)}\left[
2M_K^2(M_K^2-s_{\pi}) \right.\right. \nonumber\\
&& \left.\left.
-m_{\ell}^2(5M_K^2-3m_{\ell}^2+3s_{\pi})\right]\right\}
B(s_{\pi},m_{\ell}^2,M_K^2) -B(s_{\pi},M_{\pi}^2,M_{\pi}^2) \nonumber\\
&& -\frac{1}{2}\left\{ 9+\frac{1}{t_{\pi}}\left(
M_K^2-M_{\pi}^2-3m_{\ell}^2\right) \right. \nonumber\\
&& +\frac{8}{\lambda (t_{\pi},M_{\pi}^2,M_K^2)}\left[
(M_K^2-2M_{\pi}^2)(M_{\pi}^2-M_K^2)+(M_K^2+2M_{\pi}^2)t_{\pi}\right]
\nonumber\\
&& +\frac{m_{\ell}^2}{\lambda (t_{\pi},m_{\ell}^2,M_{\pi}^2)}\left[
2(M_K^2-13M_{\pi}^2-2m_{\ell}^2)\right. \nonumber\\
&& \left. -\frac{m_{\ell}^2}{t_{\pi}}\left(
M_K^2+5M_{\pi}^2-3m_{\ell}^2\right) +\frac{M_{\pi}^2}{t_{\pi}}\left(
M_K^2+2M_{\pi}^2\right)\right] \nonumber\\
&&
+\frac{6m_{\ell}^2M_{\pi}^2}{\lambda^2(t_{\pi},m_{\ell}^2,M_{\pi}^2)}\left[
m_{\ell}^2(M_K^2-7M_{\pi}^2+3m_{\ell}^2)\right. \nonumber\\
&& \left.\left.
-M_{\pi}^2(M_K^2-4M_{\pi}^2)+(M_K^2-4M_{\pi}^2-5m_{\ell}^2)t_{\pi}\right]\right\}
B(t_{\pi},M_{\pi}^2,M_K^2) \nonumber\\
&& +\frac{3m_{\ell}^2}{\lambda (t_{\pi},m_{\ell}^2,M_{\pi}^2)}\left(
M_{\pi}^2-m_{\ell}^2+t_{\pi}\right) B(t_{\pi},m_{\ell}^2,M_{\pi}^2)
\nonumber\\
&&
+2(M_K^2+m_{\ell}^2-s_{\pi})C(m_{\ell}^2,s_{\pi},M_K^2,m_{\gamma}^2,m_{\ell}^2,M_K^2)
\nonumber\\
&&
+2(2M_{\pi}^2-s_{\pi})C(M_{\pi}^2,s_{\pi},M_{\pi}^2,m_{\gamma}^2,M_{\pi}^2,M_{\pi}^2)
\nonumber\\
&& +\frac{2m_{\ell}^2M_K^2s_{\pi}}{\lambda
(s_{\pi},m_{\ell}^2,M_K^2)}\,C(M_K^2,s_{\pi},m_{\ell}^2,0,M_K^2,M_K^2)
\nonumber\\
&& +m_{\ell}^2\left\{ 1+\frac{1}{\lambda
(s_{\pi},m_{\ell}^2,M_K^2)}\left[
m_{\ell}^2(2M_K^2-m_{\ell}^2+s_{\pi})-M_K^2(M_K^2-s_{\pi})\right]\right.
\nonumber\\
&& +\frac{1}{\lambda (t_{\pi},m_{\ell}^2,M_{\pi}^2)}\left[
-m_{\ell}^2(M_K^2-M_{\pi}^2-m_{\ell}^2) \right.
\nonumber\\
&& \left.\left.
-M_{\pi}^2M_K^2+(M_K^2-m_{\ell}^2)t_{\pi}\right]\right\}
C(m_{\ell}^2,0,m_{\ell}^2,0,m_{\ell}^2,M_K^2) \nonumber\\
&& -m_{\ell}^2\left\{ 1-\frac{1}{\lambda
(s_{\pi},m_{\ell}^2,M_K^2)}\left[ M_K^2(M_K^2-3s_{\pi})
\right.\right. \nonumber\\
&& \left.\left.
-m_{\ell}^2(2M_K^2-m_{\ell}^2+s_{\pi})\right]\right\}
C(s_{\pi},s_{\pi},0,m_{\ell}^2,M_K^2,M_K^2) \nonumber\\
&& +\frac{m_{\ell}^2}{2}\left\{ 1+\frac{3}{t_{\pi}}\left(
m_{\ell}^2-M_{\pi}^2\right) +\frac{1}{\lambda
(t_{\pi},m_{\ell}^2,M_{\pi}^2)}\left[ -M_{\pi}^2(3M_K^2+8M_{\pi}^2)
\right.\right. \nonumber\\
&&
+m_{\ell}^2(M_K^2+7M_{\pi}^2+3m_{\ell}^2)+\frac{m_{\ell}^4}{t_{\pi}}\left(
M_K^2+8M_{\pi}^2-3m_{\ell}^2\right) \nonumber\\
&&
-2(M_K^2-3M_{\pi}^2)t_{\pi}-\frac{m_{\ell}^2M_{\pi}^2}{t_{\pi}}\left(
2M_K^2+7M_{\pi}^2\right) \nonumber\\
&& \left.\left. +\frac{M_{\pi}^4}{t_{\pi}}\left(
M_K^2+2M_{\pi}^2\right)\right]\right\}
C(t_{\pi},t_{\pi},0,m_{\ell}^2,M_{\pi}^2,M_K^2) \nonumber\\
&& -\frac{M_{\pi}^2}{\lambda (t_{\pi},m_{\ell}^2,M_{\pi}^2)}\left\{
2(M_K^2-3m_{\ell}^2)(m_{\ell}^2-M_K^2) \right. \nonumber\\
&& -\frac{3m_{\ell}^2}{\lambda (t_{\pi},m_{\ell}^2,M_{\pi}^2)}\left[
m_{\ell}^4(2M_K^2+3M_{\pi}^2-m_{\ell}^2)-2m_{\ell}^2M_{\pi}^2M_K^2
\right. \nonumber\\
&&
-M_K^4(M_{\pi}^2+m_{\ell}^2)+4M_{\pi}^4(M_K^2-m_{\ell}^2)-m_{\ell}^2(2M_K^2-4M_{\pi}^2-m_{\ell}^2)t_{\pi}
\nonumber\\
&& \left.\left. +M_K^2(M_K^2-4M_{\pi}^2)t_{\pi}\right]\right\}
C(M_{\pi}^2,t_{\pi},m_{\ell}^2,0,M_{\pi}^2,M_K^2)\,, \\
U_{e^2}^g &=& -\frac{1}{32\pi^2}\left[
9+2\ln\left(\frac{m_{\gamma}^2}{m_{\ell}^2}\right)
+4\ln\left(\frac{m_{\gamma}^2}{M_{\pi}^2}\right)
+2\ln\left(\frac{m_{\gamma}^2}{M_K^2}\right)\right] \nonumber\\
&& -\frac{1}{18}\left(
24K_1+24K_2-144K_3-72K_4+32K_5-40K_6-36K_{12}-24X_1+9X_6\right) \nonumber\\
&&
-\frac{1}{2}\,\frac{A(m_{\ell}^2)}{m_{\ell}^2}+2\,\frac{A(M_{\pi}^2)}{M_{\pi}^2}+\frac{A(M_K^2)}{M_K^2}+\frac{1}{2t_{\pi}}\left[
A(M_{\pi}^2)-A(M_K^2)\right] \nonumber\\
&& +\frac{1}{2}\left\{
3+\frac{3m_{\ell}^2}{t_{\pi}}+\frac{m_{\ell}^2}{\lambda
(t_{\pi},m_{\ell}^2,M_{\pi}^2)}\left[
\frac{m_{\ell}^2}{t_{\pi}}\left(
M_K^2+5M_{\pi}^2-3m_{\ell}^2\right)\right.\right. \nonumber\\
&& \left.\left. -\frac{M_{\pi}^2}{t_{\pi}}\left(
M_K^2+2M_{\pi}^2\right)
-(M_K^2-12M_{\pi}^2-5m_{\ell}^2+2t_{\pi})\right]\right\}
B(0,m_{\ell}^2,M_K^2) \nonumber\\
&& -\frac{1}{2}\left\{ 3-\frac{m_{\ell}^2}{\lambda
(s_{\pi},m_{\ell}^2,M_K^2)}\left( M_K^2+m_{\ell}^2-s_{\pi}\right)
\right. \nonumber\\
&& \left.
+\frac{m_{\ell}^2(M_K^2-2M_{\pi}^2+m_{\ell}^2-2t_{\pi})}{\lambda
(t_{\pi},m_{\ell}^2,M_{\pi}^2)}\left[
1+\frac{12m_{\ell}^2M_{\pi}^2}{\lambda
(t_{\pi},m_{\ell}^2,M_{\pi}^2)}\right]\right\} B(m_{\ell}^2,0,M_K^2)
\nonumber\\
&& +\left[ 2+\frac{2m_{\ell}^2}{\lambda
(s_{\pi},m_{\ell}^2,M_K^2)}\left( M_K^2-m_{\ell}^2+s_{\pi}\right)\right. \nonumber\\
&& \left. +\frac{3m_{\ell}^2}{\lambda
(t_{\pi},m_{\ell}^2,M_{\pi}^2)}\left(
M_{\pi}^2+m_{\ell}^2-t_{\pi}\right)\right]
B(m_{\ell}^2,0,m_{\ell}^2) \nonumber\\
&& +\left\{
-4+2\left(\frac{8M_{\pi}^2-3s_{\pi}}{4M_{\pi}^2-s_{\pi}}\right)+\frac{8M_{\pi}^2M_K^2}{\lambda
(t_{\pi},M_{\pi}^2,M_K^2)} \right. \nonumber\\
&&
+\frac{3m_{\ell}^2M_{\pi}^2}{\lambda^2(t_{\pi},m_{\ell}^2,M_{\pi}^2)}\left[M_{\pi}^2(M_K^2-4M_{\pi}^2)+m_{\ell}^2(M_K^2+3M_{\pi}^2-m_{\ell}^2)\right.
\nonumber\\
&& \left.\left. -(M_K^2-4M_{\pi}^2-m_{\ell}^2)t_{\pi}\right]\right\}
B(M_{\pi}^2,0,M_{\pi}^2) \nonumber\\
&& +\left[ 1-\frac{M_K^2}{\lambda (s_{\pi},m_{\ell}^2,M_K^2)}\left(
2M_K^2-m_{\ell}^2-2s_{\pi}\right)\right. \nonumber\\
&& \left. -\frac{4M_K^2}{\lambda (t_{\pi},M_{\pi}^2,M_K^2)}\left(
M_K^2+M_{\pi}^2-t_{\pi}\right)\right] B(M_K^2,0,M_K^2) \nonumber\\
&& -2\left\{ 1+\frac{1}{\lambda (s_{\pi},m_{\ell}^2,M_K^2)}\left[
m_{\ell}^2(2M_K^2-m_{\ell}^2)-M_K^4+(M_K^2+m_{\ell}^2)s_{\pi}\right]\right\}
B(s_{\pi},m_{\ell}^2,M_K^2) \nonumber\\
&& -4\left(\frac{2M_{\pi}^2-s_{\pi}}{4M_{\pi}^2-s_{\pi}}\right) B(s_{\pi},M_{\pi}^2,M_{\pi}^2)
+\frac{m_{\ell}^2(M_K^2-m_{\ell}^2+s_{\pi})}{2\lambda (s_{\pi},m_{\ell}^2,M_K^2)}\,B(s_{\pi},M_K^2,M_K^2) \nonumber\\
&& -\frac{1}{2}\left\{ 3-\frac{1}{t_{\pi}}\left(
M_K^2-M_{\pi}^2-3m_{\ell}^2\right) -\frac{8M_K^2}{\lambda
(t_{\pi},M_{\pi}^2,M_K^2)}\left(
M_K^2-M_{\pi}^2-t_{\pi}\right)\right. \nonumber\\
&& +\frac{m_{\ell}^2}{\lambda (t_{\pi},m_{\ell}^2,M_{\pi}^2)}\left[
-2(M_K^2-13M_{\pi}^2-2m_{\ell}^2)\right. \nonumber\\
&& \left. +\frac{m_{\ell}^2}{t_{\pi}}\left(
M_K^2+5M_{\pi}^2-3m_{\ell}^2\right) -\frac{M_{\pi}^2}{t_{\pi}}\left(
M_K^2+2M_{\pi}^2\right)\right] \nonumber\\
&&
-\frac{2m_{\ell}^2M_{\pi}^2}{\lambda^2(t_{\pi},m_{\ell}^2,M_{\pi}^2)}\left[
3m_{\ell}^2(M_K^2-7M_{\pi}^2+3m_{\ell}^2)\right. \nonumber\\
&& \left.\left.
-3M_{\pi}^2(M_K^2-4M_{\pi}^2)+3(M_K^2-4M_{\pi}^2-5m_{\ell}^2)t_{\pi}\right]\right\}
B(t_{\pi},M_{\pi}^2,M_K^2) \nonumber\\
&& +\frac{3m_{\ell}^2}{\lambda (t_{\pi},m_{\ell}^2,M_{\pi}^2)}\left(
M_{\pi}^2-m_{\ell}^2+t_{\pi}\right) B(t_{\pi},m_{\ell}^2,M_{\pi}^2)
\nonumber\\
&&
+2(M_K^2+m_{\ell}^2-s_{\pi})C(m_{\ell}^2,s_{\pi},M_K^2,m_{\gamma}^2,m_{\ell}^2,M_K^2)
\nonumber\\
&&
+2(2M_{\pi}^2-s_{\pi})C(M_{\pi}^2,s_{\pi},M_{\pi}^2,m_{\gamma}^2,M_{\pi}^2,M_{\pi}^2)
\nonumber\\
&& -\frac{m_{\ell}^2M_K^2}{\lambda
(s_{\pi},m_{\ell}^2,M_K^2)}\,(M_K^2-m_{\ell}^2)C(M_K^2,s_{\pi},m_{\ell}^2,0,M_K^2,M_K^2)
\nonumber\\
&& -m_{\ell}^2\left\{ 2+\frac{1}{\lambda
(t_{\pi},m_{\ell}^2,M_{\pi}^2)}\left[
m_{\ell}^2(M_K^2+7M_{\pi}^2-m_{\ell}^2) \right.\right.
\nonumber\\
&& \left.\left.
+M_{\pi}^2M_K^2-(M_K^2-m_{\ell}^2)t_{\pi}\right]\right\}
C(m_{\ell}^2,0,m_{\ell}^2,0,m_{\ell}^2,M_K^2) \nonumber\\
&& -\frac{m_{\ell}^2}{2}\,C(s_{\pi},s_{\pi},0,m_{\ell}^2,M_K^2,M_K^2) \nonumber\\
&& +\frac{m_{\ell}^2}{2}\left\{ 3-\frac{3}{t_{\pi}}\left(
m_{\ell}^2-M_{\pi}^2\right) +\frac{1}{\lambda
(t_{\pi},m_{\ell}^2,M_{\pi}^2)}\left[ -M_{\pi}^2M_K^2\right.\right. \nonumber\\
&&
+m_{\ell}^2(3M_K^2+5M_{\pi}^2-7m_{\ell}^2)-\frac{m_{\ell}^4}{t_{\pi}}\left(
M_K^2+8M_{\pi}^2-3m_{\ell}^2\right) \nonumber\\
&&
-2(M_K^2-M_{\pi}^2-2m_{\ell}^2)t_{\pi}+\frac{m_{\ell}^2M_{\pi}^2}{t_{\pi}}\left(
2M_K^2+7M_{\pi}^2\right) \nonumber\\
&& \left.\left. -\frac{M_{\pi}^4}{t_{\pi}}\left(
M_K^2+2M_{\pi}^2\right)\right]\right\}
C(t_{\pi},t_{\pi},0,m_{\ell}^2,M_{\pi}^2,M_K^2) \nonumber\\
&& +\frac{m_{\ell}^2M_{\pi}^2}{\lambda
(t_{\pi},m_{\ell}^2,M_{\pi}^2)}\left\{
4(2M_K^2+M_{\pi}^2-m_{\ell}^2-t_{\pi}) \right. \nonumber\\
&& +\frac{3}{\lambda (t_{\pi},m_{\ell}^2,M_{\pi}^2)}\left[
-m_{\ell}^4(2M_K^2+3M_{\pi}^2-m_{\ell}^2)
\right. \nonumber\\
&&
+2m_{\ell}^2M_{\pi}^2(M_K^2+2M_{\pi}^2)+M_K^4(M_{\pi}^2+m_{\ell}^2)
\nonumber\\
&& -4M_{\pi}^4M_K^2+m_{\ell}^2(2M_K^2-4M_{\pi}^2-m_{\ell}^2)t_{\pi}
\nonumber\\
&& \left.\left. -M_K^2(M_K^2-4M_{\pi}^2)t_{\pi}\right]\right\}
C(M_{\pi}^2,t_{\pi},m_{\ell}^2,0,M_{\pi}^2,M_K^2)\,.
\end{eqnarray}

\section{Results}

We shall proceed to the numerical evaluation of isospin breaking
corrections. To this end, we must handle all types of singularities
encountered in our expressions. These are of three types in general:
ultraviolet, infrared, and Coulomb. Although our expressions are
ultraviolet finite, they are infrared divergent. We showed in
Ref.~\cite{Cuplov:2003bj} that the latter singularity is canceled by
the emission of a real soft photon at the level of differential
decay rate. Since we are interested in measuring form factors, a
subtraction of infrared divergence at this level is needed. There
are infinitely many choices to do so. We shall choose the simplest
\emph{minimal subtraction scheme} consisting on dropping out, from
the expression of form factors, $\ln m_{\gamma}$ terms only.
Finally, Coulomb interaction between charged particles induces
singularities due to a photon exchange between:
\begin{enumerate}
    \item[(1)] the kaon and a pion. This occurs at $t_{\pi}=(M_K\pm
    M_{\pi})^2$ or
    \begin{equation}
    s_{\pi}\,=\,4M_{\pi}^2-(M_K\pm 2M_{\pi})^2+m_{\ell}^2\,.
    \end{equation}
    Hence, the singularity is situated outside the \emph{allowed kinematical
    region},
    \begin{equation}
    4M_{\pi}^2\,\leqslant\,s_{\pi}\,\leqslant\,(M_K-m_{\ell})^2\,,
    \end{equation}
    from the left.
    \item[(2)] the two pions. This occurs at
    \begin{equation}
    s_{\pi}\,=\,0,\,4M_{\pi}^2.
    \end{equation}
    The former value represents a pseudo-threshold and is
    situated outside the allowed kinematical region from the left.
    The latter value is a normal threshold and is situated at the
    lower bound of the allowed kinematical region. The corresponding
    singularity is of great experimental importance for the present
    work and we will study it further in the following.
    \item[(3)] the kaon and the lepton. This occurs at
    \begin{equation}
    s_{\pi}\,=\,(M_K\pm m_{\ell})^2\,.
    \end{equation}
    The pseudo-threshold is situated at the upper bound of the
    allowed kinematical region. The normal threshold outside the
    latter from the right.
    \item[(4)] a pion and the lepton. This occurs at $t_{\pi}=(M_{\pi}\pm
    m_{\ell})^2$ or
    \begin{equation}
    s_{\pi}\,=\,(M_K-m_{\ell})^2+2m_{\ell}(M_K\mp
    2M_{\pi}-m_{\ell})\,.
    \end{equation}
    Hence, the singularity is situated outside the allowed kinematical
    region from the right.
\end{enumerate}
Let us return to the Coulomb interaction between the two pions and
shift the value of $s_{\pi}$ from $4M_{\pi}^2$ by an infinitesimal
positive amount
\begin{equation}
s_{\pi}\,=4(M_{\pi}^2+\varrho^2)\,.
\end{equation}
We then expand our expressions in powers of $\varrho$. The Coulomb
singularity shows up then as poles in the $\varrho$-plane. In order
to obtain finite (regularized) results, we simply remove these poles
allowing the numerical evaluation of form factors.

\subsection{Input}

We shall use the following numerical values~\footnote{Our
expressions are evaluated at the scale $\mu$ equal to the rho mass.}
for the various parameters~\cite{Nehme:2003bz}:
\begin{enumerate}
    \item[(1)] the fine structure constant,
    \begin{equation}
    \alpha\,=\,1/137.03599976(50)\,,
    \end{equation}
    corresponding to the classical electron charge
    $e=\sqrt{4\pi\alpha}$;
    \item[(2)] the masses of the charged leptons,
    \begin{equation}
    m_e\,=\,0.510998902(21)\,\textrm{MeV}\,, \qquad
    m_{\mu}\,=\,105.658357(5)\,\textrm{MeV}\,;
    \end{equation}
    \item[(3)] the masses of the light mesons,
    \begin{eqnarray}
    && M_{\pi^{\pm}}\,=\,139.57018(35)\,\textrm{MeV}\,, \qquad
    M_{K^{\pm}}\,=\,493.677\pm 0.016\,\textrm{MeV}\,, \\
    && M_{\eta}\,=\,547.30\pm 0.12\,\textrm{MeV}\,, \qquad
    M_{\rho}\,=\,771.1\pm 0.9\,\textrm{MeV}\,;
    \end{eqnarray}
    \item[(4)] the quark masses and condensates,
    \begin{eqnarray}
    && M_{\pi}\,=\,134.9766(6)\,\textrm{MeV}\,, \quad
    M_K\,=\,495.042\pm 0.034\,\textrm{MeV}\,, \\
    && \epsilon\,=\,(1.061\pm 0.083)\times 10^{-2}\,;
    \end{eqnarray}
    \item[(5)] the low-energy constants in the strong sector,
    \begin{eqnarray}
    && L_1^r\,=\,(0.46\pm 0.24)\times 10^{-3}\,, \quad
    L_2^r\,=\,(1.49\pm 0.23)\times 10^{-3}\,, \\
    && L_3^r\,=\,(-3.18\pm 0.85)\times 10^{-3}\,, \quad
    L_4^r\,=\,(0.53\pm 0.39)\times 10^{-3}\,, \\
    && L_9^r\,=\,(5.5\pm 0.2)\times 10^{-3}\,,
    \end{eqnarray}
    \item[(6)] the low-energy constants in the electromagnetic sector,
    \begin{eqnarray}
    && K_1^r\,=\,-6.4\times 10^{-3}\,, \quad K_2^r\,=\,-3.1\times
    10^{-3}\,, \quad K_3^r\,=\,6.4\times 10^{-3}\,, \\
    && K_4^r\,=\,-6.4\times 10^{-3}\,, \quad K_5^r\,=\,19.9\times
    10^{-3}\,, \quad K_6^r\,=\,8.6\times 10^{-3}\,, \\
    && K_{12}^r\,=\,-9.2\times 10^{-3}\,,
    \end{eqnarray}
    with an error of $\pm 6.3\times 10^{-3}$ assigned to each of
    them;
    \item[(7)] the low-energy constants in the leptonic sector,
    \begin{equation}
    \mid X_i\mid\leqslant 6.3\times 10^{-3}\,;
    \end{equation}
    \item[(8)] the coupling of axial currents to the vacuum,
    \begin{equation}
    57.40\leqslant F_0\leqslant 67.53\,;
    \end{equation}
    \item[(9)] the charged pion decay constant and electromagnetic
    mass,
    \begin{equation}
    F_{\pi}\,=\,92.419\pm 0.325\,\textrm{MeV}\,, \qquad Z_0\,=0.805(1)\,.
    \end{equation}
\end{enumerate}

\subsection{Form factors}

\begin{figure}
\begin{center}
\includegraphics{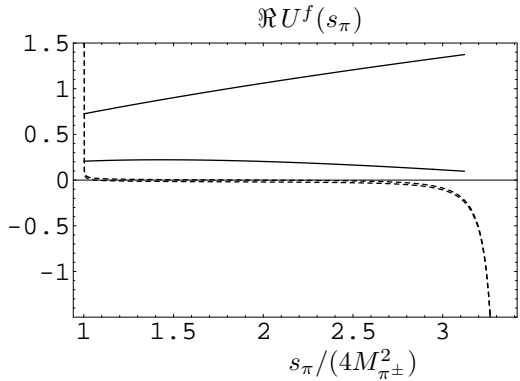}
\end{center}
\caption{\label{fig:1} Radiative correction to the real part of the
first term in the partial wave expansion for $f$ form factor under
the assumptions $s_{\ell}=m_{\ell}^2=m_e^2,\,
F_0=67.53\,\textrm{MeV}$. The plain curve represents the one-loop
correction in the absence of isospin breaking. The dashed curve
gives the isospin breaking correction of order $\mathcal{O}(\alpha
,\, m_d-m_u)$. The infrared divergence has been removed applying a
minimal subtraction scheme. Error bands come exclusively from the
uncertainty in the determination of low-energy constants and have
been developed in quadrature.}
\end{figure}

\begin{figure}
\begin{center}
\includegraphics{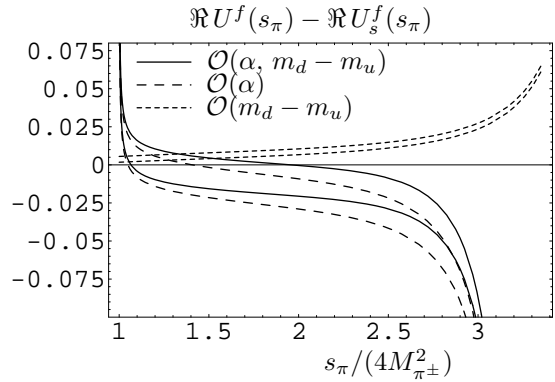}
\end{center}
\caption{\label{fig:2} Isospin breaking correction to the real part
of the first term in the partial wave expansion for $f$ form factor
under the assumptions $s_{\ell}=m_{\ell}^2=m_e^2,\,
F_0=67.53\,\textrm{MeV}$. The infrared divergence has been removed
applying a minimal subtraction scheme. Error bands come exclusively
from the uncertainty in the determination of low-energy constants
and have been developed in quadrature.}
\end{figure}

Using the preceding input parameters, we drew (Fig.~\ref{fig:1}) the
curve of the variation of one-loop level correction to the real part
for $f$ form factor as function of $s_{\pi}$. In Fig.~\ref{fig:2},
we drew the isospin breaking correction to the same quantity and
compared the contributions of orders $\mathcal{O}(\alpha )$ and
$\mathcal{O}(m_d-m_u)$.

\begin{figure}
\begin{center}
\includegraphics{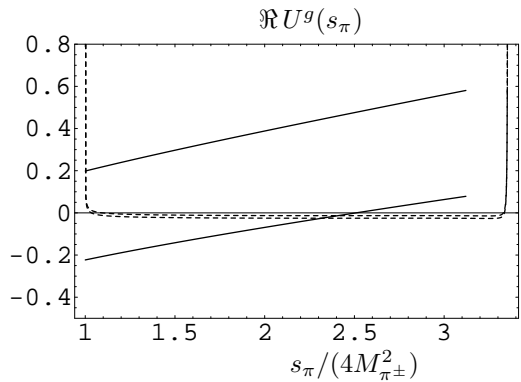}
\end{center}
\caption{\label{fig:3} Radiative correction to the real part of the
first term in the partial wave expansion for $g$ form factor under
the assumptions $s_{\ell}=m_{\ell}^2=m_e^2,\,
F_0=67.53\,\textrm{MeV}$. The plain curve represents the one-loop
correction in the absence of isospin breaking. The dashed curve
gives the isospin breaking correction of order $\mathcal{O}(\alpha
,\, m_d-m_u)$. The infrared divergence has been removed applying a
minimal subtraction scheme. Error bands come exclusively from the
uncertainty in the determination of low-energy constants and have
been developed in quadrature.}
\end{figure}

\begin{figure}
\begin{center}
\includegraphics{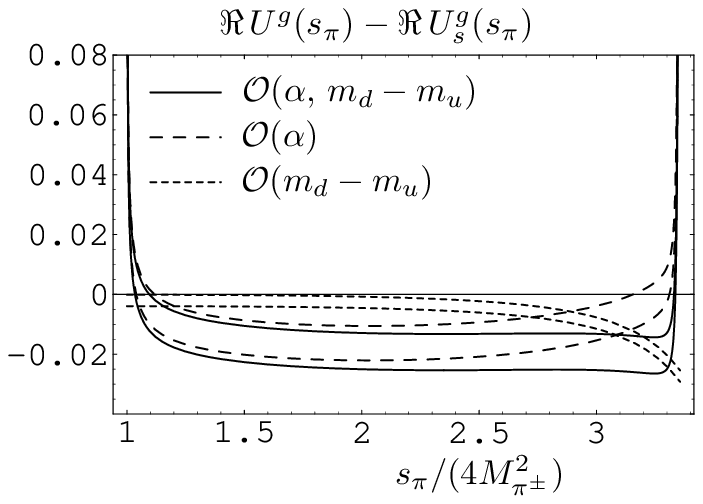}
\end{center}
\caption{\label{fig:4} Isospin breaking correction to the real part
of the first term in the partial wave expansion for $g$ form factor
under the assumptions $s_{\ell}=m_{\ell}^2=m_e^2,\,
F_0=67.53\,\textrm{MeV}$. The infrared divergence has been removed
applying a minimal subtraction scheme. Error bands come exclusively
from the uncertainty in the determination of low-energy constants
and have been developed in quadrature.}
\end{figure}

The same has been done for the real part of $g$ form factor in
Figs.~\ref{fig:3} and~\ref{fig:4}.

After removing infrared singularity and Coulomb poles, the NA48
experiment should measure what we will call \emph{subtracted form
factors}. The corresponding \emph{modules} are found to be
\begin{eqnarray}
f_S(s_{\pi}) &=& 1+\Re\,U^f(s_{\pi})+\textrm{subtraction}\,, \\
g_P(s_{\pi}) &=& 1+\Re\,U^g(s_{\pi})+\textrm{subtraction}\,, \\
\textrm{subtraction} &\doteq&
-\frac{e^2}{16}\,\frac{M_{\pi}}{\varrho}\,\textrm{If}\left(
s_{\pi}=4M_{\pi}^2\right) \nonumber\\
&& +\frac{e^2}{8\pi^2}\left[ 2+\left(
1-\frac{2M_{\pi}^2}{s_{\pi}}\right)\frac{1}{\sigma_{\pi}}\,\ln\left(\frac{1-\sigma_{\pi}}{1+\sigma_{\pi}}\right)\right]\ln
(m_{\gamma}^2) \nonumber\\
&&
+\frac{e^2}{8\pi^2}\,\frac{M_K^2+m_{\ell}^2-s_{\pi}}{\sqrt{(M_K+m_{\ell})^2-s_{\pi}}\sqrt{(M_K-m_{\ell})^2-s_{\pi}}}\,\ln
(\sigma_{\ell K})\ln (m_{\gamma}^2)\,,
\end{eqnarray}
where $\textrm{If}(argument)$ is a logical function equal to $1$ if
$argument$ is true and to $0$ if $argument$ is false.

\subsection{Phase shifts}

\begin{figure}
\begin{center}
\includegraphics{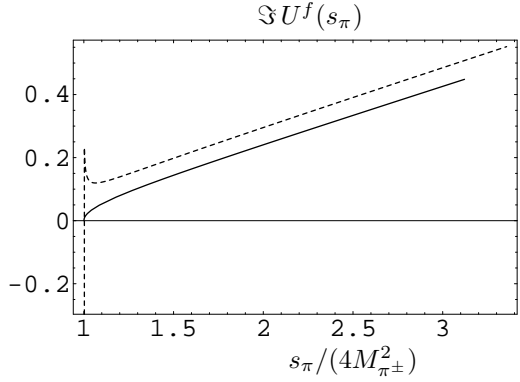}
\end{center}
\caption{\label{fig:5} The imaginary part (in radians) of the first
term in the partial wave expansion for $f$ form factor under the
assumptions $s_{\ell}=m_{\ell}^2=m_e^2,\,
F_0=F_{\pi}=92.419\,\textrm{MeV}$. The infrared divergence has been
removed applying a minimal subtraction scheme. The plain curve
represents $\delta_0^0(s_{\pi})$. The dashed one includes isospin
breaking effects.}
\end{figure}

\begin{figure}
\begin{center}
\includegraphics{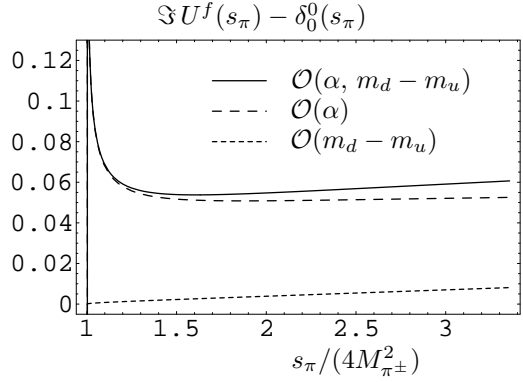}
\end{center}
\caption{\label{fig:6} Isospin breaking correction to the imaginary
part (in radians) of the first term in the partial wave expansion
for $f$ form factor under the assumptions
$s_{\ell}=m_{\ell}^2=m_e^2,\, F_0=F_{\pi}=92.419\,\textrm{MeV}$. The
infrared divergence has been removed applying a minimal subtraction
scheme.}
\end{figure}

\begin{figure}
\begin{center}
\includegraphics{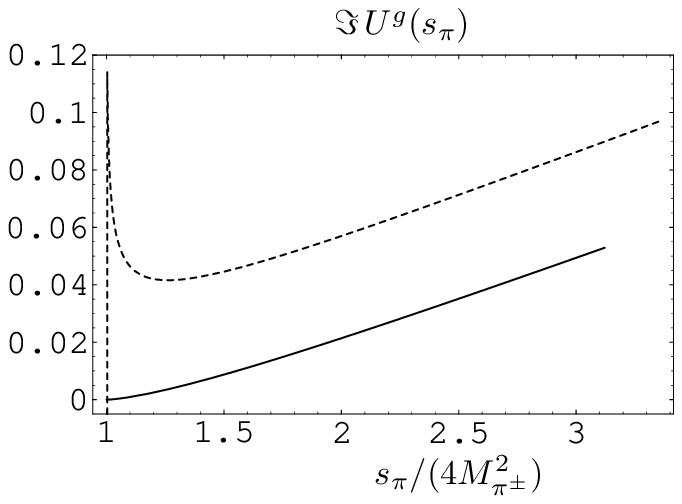}
\end{center}
\caption{\label{fig:7} The imaginary part (in radians) of the first
term in the partial wave expansion for $g$ form factor under the
assumptions $s_{\ell}=m_{\ell}^2=m_e^2,\,
F_0=F_{\pi}=92.419\,\textrm{MeV}$. The infrared divergence has been
removed applying a minimal subtraction scheme. The plain curve
represents $\delta_1^1(s_{\pi})$. The dashed one includes isospin
breaking effects.}
\end{figure}

The $S$-wave iso-scalar and $P$-wave iso-vector $\pi\pi$ phase
shifts are given by
\begin{eqnarray}
\delta_0^0(s_{\pi}) &=& \frac{1}{32\pi
F_{\pi}^2}\,(2s_{\pi}-M_{\pi^{\pm}}^2)\left(
1-\frac{4M_{\pi^{\pm}}^2}{s_{\pi}}\right)^{1/2}\,, \\
\delta_1^1(s_{\pi}) &=& \frac{1}{96\pi
F_{\pi}^2}\,(s_{\pi}-4M_{\pi^{\pm}}^2)\left(
1-\frac{4M_{\pi^{\pm}}^2}{s_{\pi}}\right)^{1/2}\,,
\end{eqnarray}
respectively. The imaginary part of the first term in the partial
wave expansion of form factors reads:
\begin{eqnarray}
\Im\,U^f(s_{\pi}) &=& \delta_0^0(s_{\pi}) \nonumber\\
&& +\frac{3}{16\pi
F_{\pi}^2}\,\frac{\epsilon}{\sqrt{3}}\,(s_{\pi}-M_{\pi}^2)\sigma_{\pi}
-\frac{\alpha}{4}\,(1-5Z_0)\sigma_{\pi} \nonumber\\
&& +\frac{\alpha}{2}\,\frac{1}{\sigma_{\pi}}\left\{\left(
1-\frac{M_{\pi}^2}{s_{\pi}}\right) Z_0+\left(
1-\frac{2M_{\pi}^2}{s_{\pi}}\right)\left[ 2\ln
(\sigma_{\pi})-\ln\left(\frac{m_{\gamma}^2}{s_{\pi}}\right)\right]\right\}
\nonumber\\
&& +\frac{3\alpha}{4}\left(
1+\frac{M_{\pi}^2-m_{\ell}^2}{t_{\pi}}\right)\frac{m_{\ell}^2}{\sqrt{t_{\pi}-(m_{\ell}+M_{\pi})^2}\sqrt{t_{\pi}-(m_{\ell}-M_{\pi})^2}}\,,
\\
\Im\,U^g(s_{\pi}) &=& \delta_1^1(s_{\pi}) \nonumber\\
&& -\frac{\alpha}{\sigma_{\pi}}\left(
1-\frac{2M_{\pi}^2}{s_{\pi}}\right)\left[ 1-\ln
(\sigma_{\pi})+\frac{1}{2}\,\ln\left(\frac{m_{\gamma}^2}{s_{\pi}}\right)\right]
\nonumber\\
&& +\frac{3\alpha}{4}\left(
1+\frac{M_{\pi}^2-m_{\ell}^2}{t_{\pi}}\right)\frac{m_{\ell}^2}{\sqrt{t_{\pi}-(m_{\ell}+M_{\pi})^2}\sqrt{t_{\pi}-(m_{\ell}-M_{\pi})^2}}\,.
\end{eqnarray}
If the infrared divergence is removed using a minimal subtraction
scheme the imaginary part for $f$ form factor takes the shape of
Fig.~\ref{fig:5}. In Fig.~\ref{fig:6}, we compared the size of each
contribution to the isospin breaking part of the same quantity.
Finally, the imaginary part for $g$ form factor is sketched in
Fig.~\ref{fig:7}. Note that the isospin breaking part is purely of
order $\mathcal{O}(\alpha )$.

For experimental purposes, we define the \emph{subtracted phase
shifts} as:
\begin{eqnarray}
\delta_S(s_{\pi}) &\doteq& \Im\,U^f(s_{\pi})
\nonumber\\
&& -\frac{\alpha}{8}\,\frac{M_{\pi}}{\varrho}\left[
3Z_0+4\ln (2\varrho )\right]\textrm{If}\left( s_{\pi}=4M_{\pi}^2\right) \nonumber\\
&& +\frac{\alpha}{2}\left(
1-\frac{2M_{\pi}^2}{s_{\pi}}\right)\frac{1}{\sigma_{\pi}}\,\ln
(m_{\gamma}^2)\,, \\
\delta_P(s_{\pi}) &\doteq& \Im\,U^g(s_{\pi}) \nonumber\\
&& +\frac{\alpha}{2}\,\frac{M_{\pi}}{\varrho}\left[ 1-\ln (2\varrho )\right]\textrm{If}\left( s_{\pi}=4M_{\pi}^2\right) \nonumber\\
&& +\frac{\alpha}{2}\left(
1-\frac{2M_{\pi}^2}{s_{\pi}}\right)\frac{1}{\sigma_{\pi}}\,\ln
(m_{\gamma}^2)\,.
\end{eqnarray}

\section{Conclusion}

In this work we made the splitting between strong and
electromagnetic interactions in $K_{\ell 4}$ decay of the charged
kaon, $K^+\rightarrow\pi^+\pi^-\ell^+\nu_{\ell}$. Our expressions
were evaluated at the production threshold for the lepton pair,
$s_{\ell}=m_{\ell}^2$. Thanks to this assumption, a partial wave
expansion of form factors with exactly the same structure as in the
pure strong theory was possible. The imaginary part of such an
expansion involves the $S$-wave isoscalar and $P$-wave isovector
$\pi\pi$ phase shifts, $\delta_0^0(s_{\pi})$ and
$\delta_1^1(s_{\pi})$, respectively. These can be related to
$\pi\pi$ scattering lengths via Roy equations. In their turn,
scattering lengths are sensitive to the way Chiral symmetry is
spontaneously broken. Consequently, a theoretical study of the
process in question including all possible contributions is
imperative. We gave here the first analytic and numerical evaluation
of the isospin breaking contribution. This would allow the
extraction of $\delta_0^0(s_{\pi})$ and $\delta_1^1(s_{\pi})$ from
the experimental measurement of form factors. Our results can be
summarized as follows:
\begin{itemize}
    \item Isospin breaking affects modules of form factors only by
    the effect of Coulomb interaction between charged particles. The
    one between the two pions is of great importance and induces a
    singularity at $s_{\pi}=4M_{\pi}^2$. We gave the residue of the
    pole in the present work.
    \item The effect of isospin breaking on the imaginary parts of form factors
    is considerable if the infrared divergence is removed using a minimal subtraction
    scheme. We gave here all analytical expressions
    for the imaginary part including the finite part, the infrared
    divergent part, the singular part with the residue of the pole.
\end{itemize}
Our results are of great utility for the interpretation of the
outgoing data from the upgraded NA48 experiment at CERN.

\appendix

\section{Loop integrals}

We use dimensional regularization and adopt the
$\overline{\textrm{MS}}$ subtraction scheme
\begin{equation}
\overline{\lambda}\,\doteq\,-\frac{1}{32\pi^2}\left[\frac{2}{4-n}+1-\gamma
+\ln (4\pi )\right]\,,
\end{equation}
where $n$ is space-time dimension and $\gamma$ the Euler constant.
All the technical material necessary for the calculation of one-loop
integrals is given in the appendix of Ref.~\cite{Nehme:2003bz}.

It is convenient to take the following notations:
\begin{eqnarray}
\sigma_P &\doteq& \sqrt{1-\frac{4M_P^2}{s_{\pi}}}\,,
\\
\sigma_{PP} &\doteq& \frac{\sigma_P-1}{\sigma_P+1}\,,
\\
\sigma_{\ell K} &\doteq&
\frac{\sqrt{(M_K+m_{\ell})^2-s_{\pi}}-\sqrt{(M_K-m_{\ell})^2-s_{\pi}}}{\sqrt{(M_K+m_{\ell})^2-s_{\pi}}+\sqrt{(M_K-m_{\ell})^2-s_{\pi}}}\,,
\\
\sigma_{\ell\pi} &\doteq&
\frac{\sqrt{t_{\pi}-(m_{\ell}+M_{\pi})^2}-\sqrt{t_{\pi}-(m_{\ell}-M_{\pi})^2}}{\sqrt{t_{\pi}-(m_{\ell}+M_{\pi})^2}+\sqrt{t_{\pi}-(m_{\ell}-M_{\pi})^2}}\,,
\\
\sigma_{\pi K} &\doteq&
\frac{\sqrt{(M_{\pi}+M_K)^2-t_{\pi}}-\sqrt{(M_{\pi}-M_K)^2-t_{\pi}}}{\sqrt{(M_{\pi}+M_K)^2-t_{\pi}}+\sqrt{(M_{\pi}-M_K)^2-t_{\pi}}}\,,
\\
\lambda^{1/2}(s_{\pi},m_{\ell}^2,M_K^2) &\doteq&
\sqrt{(m_{\ell}-M_K)^2-s_{\pi}}\sqrt{(m_{\ell}+M_K)^2-s_{\pi}}\,,
\\
\lambda^{1/2}(t_{\pi},m_{\ell}^2,M_{\pi}^2) &\doteq&
\sqrt{t_{\pi}-(m_{\ell}-M_{\pi})^2}\sqrt{t_{\pi}-(m_{\ell}+M_{\pi})^2}\,,
\\
\lambda^{1/2}(t_{\pi},M_{\pi}^2,M_K^2) &\doteq&
\sqrt{(M_{\pi}-M_K)^2-t_{\pi}}\sqrt{(M_{\pi}+M_K)^2-t_{\pi}}\,,
\\
x_0 &\doteq& \sqrt{\lambda
(t_{\pi},M_{\pi}^2,M_K^2)+4t_{\pi}(M_K^2-m_{\ell}^2)}\,,
\\
x_1 &\doteq& \lambda^{1/2}(t_{\pi},M_{\pi}^2,M_K^2)\,,
\end{eqnarray}

\subsection{$A$ integrals}

The one-point function reads:
\begin{equation}
A(m^2)\,=\,m^2\left[
-2\overline{\lambda}-\frac{1}{16\pi^2}\,\ln\left(\frac{m^2}{\mu^2}\right)\right]\,,
\end{equation}
where $\mu$ an arbitrary scale with mass dimension.

\subsection{$B$ integrals}

We need the following two-point functions:
\begin{eqnarray}
B(m_{\ell}^2,0,m_{\ell}^2) &=&
\frac{A(m_{\ell}^2)}{m_{\ell}^2}+\frac{1}{16\pi^2}\,,
\\
B(M_{\pi}^2,0,M_{\pi}^2) &=&
\frac{A(M_{\pi}^2)}{M_{\pi}^2}+\frac{1}{16\pi^2}\,,
\\
B(M_K^2,0,M_K^2) &=& \frac{A(M_K^2)}{M_K^2}+\frac{1}{16\pi^2}\,,
\\
B(0,m_{\ell}^2,M_K^2) &=&
\frac{A(m_{\ell}^2)}{m_{\ell}^2}+\frac{1}{16\pi^2}\,
\frac{M_K^2}{M_K^2-m_{\ell}^2}\,\ln\left(\frac{m_{\ell}^2}{M_K^2}\right)\,,
\\
B(m_{\ell}^2,0,M_K^2) &=&
\frac{A(M_K^2)}{M_K^2}+\frac{1}{16\pi^2}\left[ 1-\left(
1-\frac{M_K^2}{m_{\ell}^2}\right)\ln\left(
1-\frac{m_{\ell}^2}{M_K^2}\right)\right] \,,
\\
\Re\,B(s_{\pi},M_{\pi}^2,M_{\pi}^2) &=&
\frac{A(M_{\pi}^2)}{M_{\pi}^2}+\frac{1}{16\pi^2}\left[
1-\sigma_{\pi}\ln\left(\frac{1+\sigma_{\pi}}{1-\sigma_{\pi}}\right)\right]\,,
\\
\Im\,B(s_{\pi},M_{\pi}^2,M_{\pi}^2) &=&
\frac{\sigma_{\pi}}{16\pi}\,,
\\
B(s_{\pi},M_K^2,M_K^2) &=&
\frac{A(M_K^2)}{M_K^2}+\frac{1}{16\pi^2}-\frac{1}{8\pi^2}\left(\frac{4M_K^2}{s_{\pi}}-1\right)^{1/2}\arctan\left(\frac{4M_K^2}{s_{\pi}}-1\right)^{-1/2}\,,
\\
B(s_{\pi},M_{\eta}^2,M_{\eta}^2) &=&
\frac{A(M_{\eta}^2)}{M_{\eta}^2}+\frac{1}{16\pi^2}-\frac{1}{8\pi^2}\left(\frac{4M_{\eta}^2}{s_{\pi}}-1\right)^{1/2}\arctan\left(\frac{4M_{\eta}^2}{s_{\pi}}-1\right)^{-1/2}\,,
\\
B(s_{\pi},m_{\ell}^2,M_K^2) &=&
\frac{1}{2}\,\frac{A(m_{\ell}^2)}{m_{\ell}^2}+\frac{1}{2}\,\frac{A(M_K^2)}{M_K^2}+\frac{1}{16\pi^2}\left[
1-\frac{1}{2s_{\pi}}\,(m_{\ell}^2-M_K^2)\ln\left(\frac{m_{\ell}^2}{M_K^2}\right)\right]
\nonumber\\
&&
+\frac{1}{16\pi^2s_{\pi}}\,\sqrt{(m_{\ell}+M_K)^2-s_{\pi}}\sqrt{(m_{\ell}-M_K)^2-s_{\pi}}
\nonumber\\
&&
\times\ln\frac{\sqrt{(m_{\ell}+M_K)^2-s_{\pi}}+\sqrt{(m_{\ell}-M_K)^2-s_{\pi}}}{\sqrt{(m_{\ell}+M_K)^2-s_{\pi}}-\sqrt{(m_{\ell}-M_K)^2-s_{\pi}}}\,,
\\
\Re\,B(t_{\pi},m_{\ell}^2,M_{\pi}^2) &=&
\frac{1}{2}\,\frac{A(m_{\ell}^2)}{m_{\ell}^2}+\frac{1}{2}\,\frac{A(M_{\pi}^2)}{M_{\pi}^2}+\frac{1}{16\pi^2}\left[
1-\frac{1}{2t_{\pi}}\,(m_{\ell}^2-M_{\pi}^2)\ln\left(\frac{m_{\ell}^2}{M_{\pi}^2}\right)\right]
\nonumber\\
&&
-\frac{1}{16\pi^2t_{\pi}}\,\sqrt{t_{\pi}-(m_{\ell}+M_{\pi})^2}\sqrt{t_{\pi}-(m_{\ell}-M_{\pi})^2}
\nonumber\\
&&
\times\ln\frac{\sqrt{t_{\pi}-(m_{\ell}-M_{\pi})^2}+\sqrt{t_{\pi}-(m_{\ell}+M_{\pi})^2}}{\sqrt{t_{\pi}-(m_{\ell}-M_{\pi})^2}-\sqrt{t_{\pi}-(m_{\ell}+M_{\pi})^2}}\,,
\\
\Im\,B(t_{\pi},m_{\ell}^2,M_{\pi}^2) &=& \frac{1}{16\pi
t_{\pi}}\,\sqrt{t_{\pi}-(m_{\ell}+M_{\pi})^2}\sqrt{t_{\pi}-(m_{\ell}-M_{\pi})^2}\,,
\\
B(t_{\pi},M_{\pi}^2,M_K^2) &=&
\frac{1}{2}\,\frac{A(M_{\pi}^2)}{M_{\pi}^2}+\frac{1}{2}\,\frac{A(M_K^2)}{M_K^2}+\frac{1}{16\pi^2}\left[
1-\frac{1}{2t_{\pi}}\,(M_{\pi}^2-M_K^2)\ln\left(\frac{M_{\pi}^2}{M_K^2}\right)\right]
\nonumber\\
&&
+\frac{1}{16\pi^2t_{\pi}}\,\sqrt{(M_{\pi}+M_K)^2-t_{\pi}}\sqrt{(M_{\pi}-M_K)^2-t_{\pi}}
\nonumber\\
&&
\times\ln\frac{\sqrt{(M_{\pi}+M_K)^2-t_{\pi}}+\sqrt{(M_{\pi}-M_K)^2-t_{\pi}}}{\sqrt{(M_{\pi}+M_K)^2-t_{\pi}}-\sqrt{(M_{\pi}-M_K)^2-t_{\pi}}}\,,
\\
B(t_{\pi},M_{\eta}^2,M_K^2) &=&
\frac{1}{2}\,\frac{A(M_{\eta}^2)}{M_{\eta}^2}+\frac{1}{2}\,\frac{A(M_K^2)}{M_K^2}+\frac{1}{16\pi^2}\left[
1-\frac{1}{2t_{\pi}}\,(M_{\eta}^2-M_K^2)\ln\left(\frac{M_{\eta}^2}{M_K^2}\right)\right]
\nonumber\\
&&
-\frac{1}{8\pi^2t_{\pi}}\,\sqrt{(M_{\eta}+M_K)^2-t_{\pi}}\sqrt{t_{\pi}-(M_{\eta}-M_K)^2}
\nonumber\\
&&
\times\arctan\frac{\sqrt{t_{\pi}-(M_{\eta}-M_K)^2}}{\sqrt{(M_{\eta}+M_K)^2-t_{\pi}}}\,,
\end{eqnarray}
For the following integral, we shall distinguish between two cases.
\begin{itemize}
    \item[(a)] The lepton is an electron:
    \begin{eqnarray}
    B(s_{\pi},M_{\pi}^2,M_{\eta}^2)
    &=&
    \frac{1}{2}\,\frac{A(M_{\pi}^2)}{M_{\pi}^2}+\frac{1}{2}\,\frac{A(M_{\eta}^2)}{M_{\eta}^2}+\frac{1}{16\pi^2}\left[
    1-\frac{1}{2s_{\pi}}\,(M_{\eta}^2-M_{\pi}^2)\ln\left(\frac{M_{\eta}^2}{M_{\pi}^2}\right)\right]
    \nonumber\\
    && +\textrm{If}\left(
    4M_{\pi}^2<s_{\pi}<(M_{\eta}-M_{\pi})^2\right) \nonumber\\
    && \times\frac{1}{16\pi^2s_{\pi}}\,\sqrt{(M_{\eta}+M_{\pi})^2-s_{\pi}}\sqrt{(M_{\eta}-M_{\pi})^2-s_{\pi}}
    \nonumber\\
    &&
    \times\ln\frac{\sqrt{(M_{\eta}+M_{\pi})^2-s_{\pi}}+\sqrt{(M_{\eta}-M_{\pi})^2-s_{\pi}}}{\sqrt{(M_{\eta}+M_{\pi})^2-s_{\pi}}-\sqrt{(M_{\eta}-M_{\pi})^2-s_{\pi}}} \nonumber\\
    && -\textrm{If}\left(
    (M_{\eta}-M_{\pi})^2<s_{\pi}<(M_K-m_e)^2\right) \nonumber\\
    &&
    \times\frac{1}{8\pi^2s_{\pi}}\,\sqrt{(M_{\eta}+M_{\pi})^2-s_{\pi}}\sqrt{s_{\pi}-(M_{\eta}-M_{\pi})^2} \nonumber\\
    && \times\arctan\frac{\sqrt{s_{\pi}-(M_{\eta}-M_{\pi})^2}}{\sqrt{(M_{\eta}+M_{\pi})^2-s_{\pi}}}\,,
    \end{eqnarray}
    \item[(b)] The lepton is a muon:
    \begin{eqnarray}
    B(s_{\pi},M_{\pi}^2,M_{\eta}^2)
    &=&\frac{1}{2}\,\frac{A(M_{\pi}^2)}{M_{\pi}^2}+\frac{1}{2}\,\frac{A(M_{\eta}^2)}{M_{\eta}^2}+\frac{1}{16\pi^2}\left[
    1-\frac{1}{2s_{\pi}}\,(M_{\eta}^2-M_{\pi}^2)\ln\left(\frac{M_{\eta}^2}{M_{\pi}^2}\right)\right]
    \nonumber\\
    &&
    +\frac{1}{16\pi^2s_{\pi}}\,\sqrt{(M_{\eta}+M_{\pi})^2-s_{\pi}}\sqrt{(M_{\eta}-M_{\pi})^2-s_{\pi}}
    \nonumber\\
    && \times\ln\frac{\sqrt{(M_{\eta}+M_{\pi})^2-s_{\pi}}+\sqrt{(M_{\eta}-M_{\pi})^2-s_{\pi}}}{\sqrt{(M_{\eta}+M_{\pi})^2-s_{\pi}}-\sqrt{(M_{\eta}-M_{\pi})^2-s_{\pi}}}
    \,.
    \end{eqnarray}
\end{itemize}

\subsection{$\tau$ integrals}

These integrals appeared while splitting strong and electromagnetic
parts in two-point functions. We are interested in the following
particular $\tau$ integrals:
\begin{eqnarray}
\Re\,\tau (s_{\pi},M_{\pi}^2,M_{\pi}^2) &=&
-\frac{2}{s_{\pi}\sigma_{\pi}}\,\ln\left(\frac{1+\sigma_{\pi}}{1-\sigma_{\pi}}\right)\,,
\\
\Im\,\tau (s_{\pi},M_{\pi}^2,M_{\pi}^2) &=&
\frac{2\pi}{s_{\pi}\sigma_{\pi}}\,,
\\
\tau (s_{\pi},M_K^2,M_K^2) &=&
\frac{4}{s_{\pi}}\left(\frac{4M_K^2}{s_{\pi}}-1\right)^{-1/2}\arctan\left(\frac{4M_K^2}{s_{\pi}}-1\right)^{-1/2}\,,
\\
\tau (t_{\pi},M_{\pi}^2,M_K^2) &=&
\frac{2}{\sqrt{(M_{\pi}-M_K)^2-t_{\pi}}\sqrt{(M_{\pi}+M_K)^2-t_{\pi}}}
\nonumber\\
&&
\times\ln\frac{\sqrt{(M_{\pi}+M_K)^2-t_{\pi}}+\sqrt{(M_{\pi}-M_K)^2-t_{\pi}}}{\sqrt{(M_{\pi}+M_K)^2-t_{\pi}}-\sqrt{(M_{\pi}-M_K)^2-t_{\pi}}}\,,
\\
\tau (t_{\pi},M_{\eta}^2,M_K^2) &=&
\frac{4}{\sqrt{t_{\pi}-(M_{\eta}-M_K)^2}\sqrt{(M_{\eta}+M_K)^2-t_{\pi}}}
\nonumber\\
&&
\times\arctan\frac{\sqrt{t_{\pi}-(M_{\eta}-M_K)^2}}{\sqrt{(M_{\eta}+M_K)^2-t_{\pi}}}\,.
\end{eqnarray}

\subsection{$C$ integrals}

These are scalar three-point functions whose definition and
expressions were given in the appendix of Ref.~\cite{Nehme:2003bz}.
In what follows, we sketch some of the particular cases that we need
for the numerical evaluation of isospin breaking corrections:
\begin{eqnarray}
C(m_{\ell}^2,0,m_{\ell}^2,0,m_{\ell}^2,M_K^2) &=&
\frac{1}{16\pi^2}\left[\frac{1}{m_{\ell}^2}\,\ln\left(
1-\frac{m_{\ell}^2}{M_K^2}\right)+\frac{1}{M_K^2-m_{\ell}^2}\,\ln\left(\frac{m_{\ell}^2}{M_K^2}\right)\right]\,,
\\
C(s_{\pi},s_{\pi},0,m_{\ell}^2,M_K^2,M_K^2) &=&
\frac{1}{M_K^2-m_{\ell}^2}\left[
B(s_{\pi},M_K^2,M_K^2)-B(s_{\pi},m_{\ell}^2,M_K^2)\right] \,,
\\
\Re\,C(M_{\pi}^2,s_{\pi},M_{\pi}^2,m_{\gamma}^2,M_{\pi}^2,M_{\pi}^2)
&=& -\frac{1}{32\pi^2s_{\pi}\sigma_{\pi}}\left\{
4\textrm{Li}_2\left(\frac{1-\sigma_{\pi}}{1+\sigma_{\pi}}\right)
+\frac{4\pi^2}{3} \right. \nonumber\\
&& \left. +\left[ 4\ln
(\sigma_{\pi})-2\ln\left(\frac{m_{\gamma}^2}{s_{\pi}}\right)
+\ln\left(\frac{1-\sigma_{\pi}}{1+\sigma_{\pi}}\right)\right]\ln\left(\frac{1-\sigma_{\pi}}{1+\sigma_{\pi}}\right)\right\}
\,,
\\
\Im\,C(M_{\pi}^2,s_{\pi},M_{\pi}^2,m_{\gamma}^2,M_{\pi}^2,M_{\pi}^2)
&=& -\frac{1}{16\pi s_{\pi}\sigma_{\pi}}\left[ 2\ln
(\sigma_{\pi})-\ln\left(\frac{m_{\gamma}^2}{s_{\pi}}\right)\right]\,,
\\
C(m_{\ell}^2,s_{\pi},M_K^2,m_{\gamma}^2,m_{\ell}^2,M_K^2) &=&
\frac{1}{16\pi^2}\,\frac{1}{M_Km_{\ell}}\,\frac{\sigma_{\ell K}}{1-\sigma_{\ell K}^2} \nonumber\\
&& \times\left\{\left[ 2\ln (1-\sigma_{\ell K}^2)-\frac{1}{2}\,\ln
(\sigma_{\ell
K})-\ln\left(\frac{m_{\gamma}^2}{M_Km_{\ell}}\right)\right]\ln
(\sigma_{\ell K}) \right.\nonumber\\
&&
-\frac{\pi^2}{6}+\frac{1}{2}\,\ln^2\left(\frac{m_{\ell}}{M_K}\right)
+\textrm{Li}_2(\sigma_{\ell
K}^2)\nonumber\\
&& \left. +\textrm{Li}_2\left( 1-\frac{m_{\ell}}{M_K}\,\sigma_{\ell
K}\right) +\textrm{Li}_2\left( 1-\frac{M_K}{m_{\ell}}\,\sigma_{\ell
K}\right)\right\}\,,
\\
C(M_K^2,s_{\pi},m_{\ell}^2,0,M_K^2,M_K^2) &=&
\frac{1}{16\pi^2}\,\frac{1}{M_Km_{\ell}}\,\frac{\sigma_{\ell
K}}{1-\sigma_{\ell K}^2} \nonumber\\
&& \times\left[\ln\left(\frac{M_K^2}{M_K^2-m_{\ell}^2}\right)
+\ln\left(\frac{M_Km_{\ell}}{M_K^2-m_{\ell}^2}\right) \right.
\nonumber\\
&&
-\ln^2(\sigma_{KK})-\frac{1}{2}\,\ln^2\left(\frac{m_{\ell}}{M_K}\right)
-\frac{1}{2}\,\ln^2(\sigma_{\ell K}) \nonumber\\
&& +\textrm{Li}_2\left( 1-\frac{m_{\ell}}{M_K}\,\sigma_{\ell
K}\right) +\textrm{Li}_2\left( 1-\frac{M_K}{m_{\ell}}\,\sigma_{\ell
K}\right) \nonumber\\
&& -\textrm{Li}_2\left( 1-\frac{M_K}{m_{\ell}}\,\frac{\sigma_{\ell
K}}{\sigma_{KK}}\right) -\textrm{Li}_2\left(
1-\frac{M_K}{m_{\ell}}\,\sigma_{\ell K}\sigma_{KK}\right)
\nonumber\\
&& \left. -\textrm{Li}_2\left(
1-\frac{m_{\ell}}{M_K}\,\frac{\sigma_{\ell K}}{\sigma_{KK}}\right)
-\textrm{Li}_2\left( 1-\frac{m_{\ell}}{M_K}\,\sigma_{\ell
K}\sigma_{KK}\right)\right\}\,,
\\
C(t_{\pi},t_{\pi},0,m_{\ell}^2,M_{\pi}^2,M_K^2)
&=& \frac{1}{32\pi^2t_{\pi}}\,\frac{1}{M_K^2-m_{\ell}^2}\left\{
(M_K^2-M_{\pi}^2+t_{\pi})\ln\left(\frac{m_{\ell}^2}{M_K^2}\right)
\right. \nonumber\\
&&
+x_0\ln\frac{M_K^2-M_{\pi}^2+t_{\pi}+x_0}{M_K^2-M_{\pi}^2+t_{\pi}-x_0}-x_1\ln\frac{M_K^2-M_{\pi}^2+t_{\pi}+x_1}{M_K^2-M_{\pi}^2+t_{\pi}-x_1}
\nonumber\\
&& -x_0\ln\frac{(x_0+M_K^2-m_{\ell}^2)^2-\lambda
(t_{\pi},m_{\ell}^2,M_{\pi}^2)}{(x_0-M_K^2+m_{\ell}^2)^2-\lambda
(t_{\pi},m_{\ell}^2,M_{\pi}^2)} \nonumber\\
&& +x_1\ln\frac{(x_1+M_K^2-m_{\ell}^2)^2-\lambda
(t_{\pi},m_{\ell}^2,M_{\pi}^2)}{(x_1-M_K^2+m_{\ell}^2)^2-\lambda
(t_{\pi},m_{\ell}^2,M_{\pi}^2)} \nonumber\\
&& -(M_K^2-m_{\ell}^2)\ln\frac{(x_0+M_K^2-m_{\ell}^2)^2-\lambda
(t_{\pi},m_{\ell}^2,M_{\pi}^2)}{(x_1+M_K^2-m_{\ell}^2)^2-\lambda
(t_{\pi},m_{\ell}^2,M_{\pi}^2)} \nonumber\\
&& -(M_K^2-m_{\ell}^2)\ln\frac{(x_0-M_K^2+m_{\ell}^2)^2-\lambda
(t_{\pi},m_{\ell}^2,M_{\pi}^2)}{(x_1-M_K^2+m_{\ell}^2)^2-\lambda
(t_{\pi},m_{\ell}^2,M_{\pi}^2)} \nonumber\\
&&
-\lambda^{1/2}(t_{\pi},m_{\ell}^2,M_{\pi}^2)\ln\frac{M_K^2-m_{\ell}^2+x_0+\lambda^{1/2}(t_{\pi},m_{\ell}^2,M_{\pi}^2)}{M_K^2-m_{\ell}^2+x_0-\lambda^{1/2}(t_{\pi},m_{\ell}^2,M_{\pi}^2)}
\nonumber\\
&&
-\lambda^{1/2}(t_{\pi},m_{\ell}^2,M_{\pi}^2)\ln\frac{M_K^2-m_{\ell}^2-x_0+\lambda^{1/2}(t_{\pi},m_{\ell}^2,M_{\pi}^2)}{M_K^2-m_{\ell}^2-x_0-\lambda^{1/2}(t_{\pi},m_{\ell}^2,M_{\pi}^2)}
\nonumber\\
&&
+\lambda^{1/2}(t_{\pi},m_{\ell}^2,M_{\pi}^2)\ln\frac{M_K^2-m_{\ell}^2-x_1+\lambda^{1/2}(t_{\pi},m_{\ell}^2,M_{\pi}^2)}{M_K^2-m_{\ell}^2-x_1-\lambda^{1/2}(t_{\pi},m_{\ell}^2,M_{\pi}^2)}
\nonumber\\
&& \left.
+\lambda^{1/2}(t_{\pi},m_{\ell}^2,M_{\pi}^2)\ln\frac{M_K^2-m_{\ell}^2+x_1+\lambda^{1/2}(t_{\pi},m_{\ell}^2,M_{\pi}^2)}{M_K^2-m_{\ell}^2+x_1-\lambda^{1/2}(t_{\pi},m_{\ell}^2,M_{\pi}^2)}\right\}\,,
\\
C(M_{\pi}^2,t_{\pi},m_{\ell}^2,0,M_{\pi}^2,M_K^2) &=&
\frac{1}{16\pi^2}\,\frac{1}{m_{\ell}M_{\pi}}\,\frac{\sigma_{\ell\pi}}{1-\sigma_{\ell\pi}^2}
\nonumber\\
&& \times\left\{\ln
(-\sigma_{\ell\pi})\left[\ln\left(\frac{m_{\ell}M_K}{M_K^2-m_{\ell}^2}\right)+\ln\left(\frac{M_{\pi}M_K}{M_K^2-m_{\ell}^2}\right)\right]
-\frac{\pi^2}{6} \right. \nonumber\\
&& +\frac{1}{2}\,\ln^2\left(\frac{m_{\ell}}{M_{\pi}}\right)
-\ln^2\left(\frac{m_{\ell}}{M_K}\right)
-\frac{1}{2}\,\ln^2(-\sigma_{\ell\pi})-\ln^2(\sigma_{\pi K})
\nonumber\\
&& -\frac{1}{2}\,\ln^2\left(
1-\frac{m_{\ell}}{M_{\pi}}\,\sigma_{\ell\pi}\right)-\frac{1}{2}\,\ln^2\left(
1-\frac{M_{\pi}}{m_{\ell}}\,\sigma_{\ell\pi}\right)
\nonumber\\
&& +\frac{1}{2}\,\ln^2\left(
1-\frac{m_{\ell}}{M_K}\,\frac{\sigma_{\ell\pi}}{\sigma_{\pi
K}}\right) +\frac{1}{2}\,\ln^2\left(1
-\frac{M_K}{m_{\ell}}\,\frac{\sigma_{\ell\pi}}{\sigma_{\pi
K}}\right) \nonumber\\
&& +\frac{1}{2}\,\ln^2\left(
1-\frac{m_{\ell}}{M_K}\,\sigma_{\ell\pi}\sigma_{\pi K}\right)
+\frac{1}{2}\,\ln^2\left(
1-\frac{M_K}{m_{\ell}}\,\sigma_{\ell\pi}\sigma_{\pi K}\right)
\nonumber\\
&&
-\textrm{Li}_2\left(\frac{m_{\ell}}{m_{\ell}-M_{\pi}\sigma_{\ell\pi}}\right)
-\textrm{Li}_2\left(\frac{M_{\pi}}{M_{\pi}-m_{\ell}\sigma_{\ell\pi}}\right)
\nonumber\\
&&
+\textrm{Li}_2\left(\frac{m_{\ell}}{m_{\ell}-M_K\sigma_{\ell\pi}\sigma_{\pi
K}}\right)
+\textrm{Li}_2\left(\frac{M_K}{M_K-m_{\ell}\sigma_{\ell\pi}\sigma_{\pi
K}}\right) \nonumber\\
&& \left. +\textrm{Li}_2\left(\frac{m_{\ell}\sigma_{\pi
K}}{m_{\ell}\sigma_{\pi K}-M_K\sigma_{\ell\pi}}\right)
+\textrm{Li}_2\left(\frac{M_K\sigma_{\pi K}}{M_K\sigma_{\pi
K}-m_{\ell}\sigma_{\ell\pi}}\right)\right\}\,,
\end{eqnarray}

\end{document}